\documentclass[sigplan,nonacm]{acmart}
\setcopyright{none}              % don't try to insert CC/ACM badges/logos
\acmConference[]{}{}{}           % clear conference fields
\acmBooktitle{}
\acmPrice{}
\acmISBN{}
\acmDOI{}

\usepackage{booktabs}
\usepackage{amsmath,amsfonts}
\usepackage{algorithm}
\usepackage{algpseudocode}
\usepackage[acronym]{glossaries}
\makeglossaries
\usepackage{multirow}
\usepackage{tikz}
\usepackage{stfloats} 
\usetikzlibrary{patterns}
\usetikzlibrary{patterns.meta}
\usepackage{soul}
\usepackage[backgroundcolor=yellow]{mdframed}

\algrenewcommand\algorithmicfunction{\textbf{Function}}
\algrenewcommand\algorithmicreturn{\textbf{return}}
\algrenewcommand\algorithmicif{\textbf{if}}
\algrenewcommand\algorithmicthen{\textbf{then}}
\algrenewcommand\algorithmicelse{\textbf{else}}
\algrenewcommand\algorithmicwhile{\textbf{while}}
\algrenewcommand\algorithmicfor{\textbf{for}}
\algrenewcommand\algorithmicdo{\textbf{do}}
\algrenewcommand\algorithmicindent{1em}

\usepackage{graphicx}
\usepackage{textcomp}
\usepackage{xspace}
\usepackage{siunitx}
\usepackage{pgf}
\usepackage{makecell}

\everymath=\expandafter{\the\everymath\displaystyle}
\usepackage{subcaption}
\usepackage{array}
\newcolumntype{C}[1]{>{\centering\arraybackslash}m{#1}}
\usepackage{xcolor}
\usepackage{colortbl}
\usepackage[most]{tcolorbox}
\usepackage{pifont}
\definecolor{darkgreen}{RGB}{0,165,0}
\newcommand{\cmark}{\textcolor{darkgreen}{\checkmark}}
\usepackage{xcolor} 
\newcommand{\cross}{\textcolor{red!100!black}{\ding{55}}}

\newtcolorbox{insightbox}[1][]{
  enhanced,
  colback=gray!5!white,
  colframe=gray!50!black,
  boxrule=0.5pt,
  left=4mm,
  right=4mm,
  top=2mm,
  bottom=2mm,
  fonttitle=\bfseries\sffamily,
  coltitle=black,
  toptitle=1mm,
  bottomtitle=1mm,
  #1
}

\newcommand{\tcompute}{T_{\text{cmp}}}

\usepackage{adjustbox}

\newacronym{edp}{EDP}{energy-delay product}
\newacronym{edpc}{EDP$\times$\$}{energy-delay-cost product}
\newacronym{ec}{energy$\times$\$}{energy-cost product}
\newacronym{cnn}{CNN}{convolutional neural network}
\newacronym{nre}{NRE}{Non-Recurring Engineering}
\newacronym{re}{RE}{Recurring Engineering}
\newacronym{vit}{VT}{vision transformer}
\newacronym{cgra}{CGRA}{coarse-grained reconfigurable array}
\newacronym{tp}{TP}{tensor parallelism}
\newacronym{tf}{TF}{tensor fusion}
\newacronym{replk}{ReplkNet31b}{ReplkNet31b}
\newacronym{db}{DB}{double buffering}
\newacronym{mobilenet}{Mobilenet}{Mobilenet v3 Small}
\newacronym{swap}{SWaP}{size, weight, and power}
\newacronym{ttft}{TTFT}{time-to-first-token}
\newacronym{mcm}{MCM}{multi-chip module}
\newacronym{tar}{TAR}{token acceptance rate}
\newacronym{bao}{BAO}{batch-agnostic operator}
\newacronym{bso}{BSO}{batch-sensitive operator}
\newacronym{nsic}{BASIC}{bespoke application-specific integrated circuit}
\newacronym{tco}{TCO}{Total Cost of Ownership}

\begin{document}
\title{Mozart: A Chiplet Ecosystem-Accelerator Codesign Framework for Composable Bespoke Application Specific Integrated Circuits}

\author{Haoran Jin}
\affiliation{
  \department{Computer Science \& Engineering}
  \institution{University of Michigan}
  \city{Ann Arbor}
  \state{MI}
  \country{USA}
}
\email{allenjin@umich.edu}

\author{Jirong Yang}
\affiliation{
  \department{Computer Science \& Engineering}
  \institution{University of Michigan}
  \city{Ann Arbor}
  \state{MI}
  \country{USA}
}
\email{yjrcs@umich.edu}

\author{Yunpeng Liu}
\affiliation{
  \department{Electrical \& Computer Engineering}
  \institution{University of Michigan}
  \city{Ann Arbor}
  \state{MI}
  \country{USA}
}
\email{yunpengl@umich.edu}

\author{Barry Lyu}
\affiliation{
  \department{Electrical \& Computer Engineering}
  \institution{University of Michigan}
  \city{Ann Arbor}
  \state{MI}
  \country{USA}
}
\email{barrylyu@umich.edu}

\author{Kangqi Zhang}
\affiliation{
  \department{Computer Science \& Engineering}
  \institution{University of Michigan}
  \city{Ann Arbor}
  \state{MI}
  \country{USA}
}
\email{zhkangqi@umich.edu}

\author{Nathaniel Bleier}
\affiliation{
  \department{Computer Science \& Engineering}
  \institution{University of Michigan}
  \city{Ann Arbor}
  \state{MI}
  \country{USA}
}
\email{nbleier@umich.edu}

\begin{abstract}
Modern AI acceleration faces a fundamental challenge: conventional
assumptions about memory requirements, batching effectiveness, and
latency-throughput tradeoffs are system-wide generalizations that ignore the
heterogeneous computational patterns of individual neural network operators.
This operator-level analysis reveals that architectural solutions must
operate at the granularity of specific computational patterns rather than
entire networks.
However, these network-level customization and operator-level heterogeneity incur substantial \gls{nre} costs.
While chiplet-based approaches have been proposed to amortize \gls{nre} costs, reuse opportunities remain limited without carefully identifying which chiplets are truly necessary.
This paper introduces Mozart, a chiplet ecosystem and accelerator
codesign framework that systematically constructs low cost \glspl{nsic}.
\Glspl{nsic} are constructed using operator-level disaggregation insights, exploring
chiplet and memory heterogeneity, tensor fusion, and tensor parallelism decisions.
The hierarchical design space exploration incorporates
novel algorithmic optimizations and integrated place-and-route validation
for efficiency and physical
implementability. 
The framework also enables constraint-aware system-level optimization across
deployment contexts ranging from datacenter inference serving to
edge computing in autonomous vehicles.

The evaluation confirms that with just 8 strategically selected
chiplet,
Mozart-generated composite \glspl{nsic}
achieve 43.5\%, 25.4\%, 67.7\%, and 78.8\%
reductions in energy, \gls{ec}, \gls{edp}, and \gls{edpc} compared to traditional
homogeneous accelerators while maintaining performance within 91-95\%
of unconstrained heterogeneous \glspl{nsic} designs across a wide range of neural networks.

For datacenter LLM serving, Mozart achieves 15-19\% energy
reduction and 35-39\% energy-cost improvement. 
In speculative decoding, Mozart delivers
throughput improvements of 24.6-58.6\% while reducing energy consumption
by 38.6-45.6\%. 
For autonomous vehicle perception, Mozart reduces energy×cost by
25.54\% and energy by 10.53\% under real-time constraints.

\end{abstract}

\maketitle % should come after the abstract

\begin{figure}[!t]
    \centering
    \includegraphics[width=0.95\columnwidth]{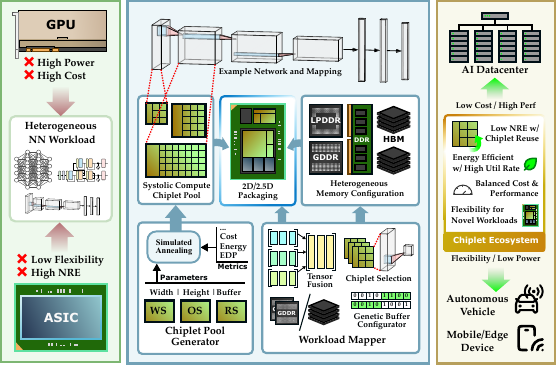}
    \caption{\label{fig:overview}
    Ideal neural network accelerators
    can support heterogeneous workloads, while
    still being flexible enough to support emerging
    workloads. They can be deployed to support
    various resource constrained applications.
    They can be designed and manufactured at low cost.}
\end{figure}

\section{Introduction}
\label{sec:introduction}

Modern AI applications span diverse domains—from datacenter inference serving to autonomous vehicle perception, from classification to generative modeling~\cite{brown2020languagemodelsfewshotlearners, vit, stable_diffusion, driving}—creating heterogeneity at three distinct levels.

First, inter-network heterogeneity emerges from the dramatic diversification of neural network architectures. 
The continuously evolving ecosystem includes specialized architectures, including attention-based transformers for sequence modeling~\cite{vit, vaswani2023attentionneed}, convolutional neural networks for spatial feature extraction~\cite{alexnet}, and generative models for content synthesis~\cite{stable_diffusion, gan}.

Second, intra-network heterogeneity arises from vastly different computational patterns within single networks. 
Modern architectures deliberately combine disparate operations to achieve superior performance. For instance, RepLKNet~\cite{Scaling_Up_Your_Kernels_to_31x31} interleaves large (31×31) and small (3×3) convolution kernels to balance receptive field and computational efficiency. 
Similarly, transformers~\cite{vaswani2023attentionneed} orchestrate element-wise operations, matrix multiplications, and attention mechanisms, each with distinct computational characteristics.

Third, inter-application heterogeneity stems from divergent deployment requirements across use cases. Latency constraints vary dramatically—from 2.5 seconds for chatbot interactions to 15 seconds for document summarization~\cite{distservedisaggregatingprefilldecoding}. Simultaneously, energy efficiency and \gls{tco} have become paramount concerns driven by power infrastructure limitations and environmental sustainability requirements~\cite{41606, wu2022sustainableaienvironmentalimplications}.

These three levels of heterogeneity expose two fundamental limitations in traditional accelerator design: lack of network-level customization and absence of operator-level heterogeneity.

Existing accelerators fail to customize for individual networks. General-purpose accelerators like GPUs optimize for broad parallelism patterns~\cite{gpuenergy, silvano2025surveydeeplearninghardware}, while domain-specific accelerators like Eyeriss target entire network families (e.g., all CNNs) rather than specific networks~\cite{eyeriss, eyeriss_v2}. 
As shown in Table~\ref{tab:intranet}, accelerators optimized for one network exhibit significant performance degradation when executing others.

\begin{table}[b]
\centering
\caption{\label{tab:intranet}
    Inter-network accelerator performance comparison}
    \begin{adjustbox}{width=\linewidth}
\begin{tabular}{|l|c|c|c|c|c|}
\hline
\multirow{2}{*}{\textbf{Network}} & \multicolumn{5}{c|}{\textbf{Accelerator Optimized for}} \\
\cline{2-6}
 & \textbf{replknet31b} & \textbf{resnet50} & \textbf{OPT-66B\_prefill B1} & \textbf{OPT-66B\_decode} & \textbf{OPT-66B\_prefill B4} \\
\hline
\hline
\textbf{replknet31b} & \cellcolor{gray!25}1.00, 1.00 & \cellcolor{yellow!15}0.93, 0.90 & \cellcolor{red!25}2.05, 3.00 & \cellcolor{yellow!25}1.07, \cellcolor{yellow!15}0.85 & \cellcolor{red!25}2.86, 2.28 \\
\hline
\textbf{resnet50} & \cellcolor{orange!25}1.50, 1.22 & \cellcolor{gray!25}1.00, 1.00 & \cellcolor{red!25}2.47, 1.97 & \cellcolor{yellow!25}1.17, \cellcolor{yellow!15}1.04 & \cellcolor{red!25}2.70, 2.24 \\
\hline
\textbf{OPT-66B\_prefill B1} & \cellcolor{red!25}2.07, \cellcolor{red!45}26.96 & \cellcolor{red!25}2.10, \cellcolor{red!40}8.25 & \cellcolor{gray!25}1.00, 1.00 & \cellcolor{red!25}1.98, \cellcolor{red!45}23.37 & \cellcolor{yellow!15}1.02, 0.88 \\
\hline
\textbf{OPT-66B\_decode B1} & \cellcolor{yellow!15}1.01, \cellcolor{yellow!25}1.05 & \cellcolor{gray!25}1.00, 1.00 & \cellcolor{yellow!15}1.03, 1.02 & \cellcolor{gray!25}1.00, 1.00 & \cellcolor{yellow!15}1.02, 1.03 \\
\hline
\textbf{OPT-66B\_prefill B4} & \cellcolor{red!25}2.70, \cellcolor{red!45}41.06 & \cellcolor{red!25}2.78, \cellcolor{red!45}13.35 & \cellcolor{yellow!15}0.99, 1.04 & \cellcolor{red!25}2.55, \cellcolor{red!45}41.71 & \cellcolor{gray!25}1.00, 1.00 \\
\hline
\end{tabular}
\end{adjustbox}
\begin{flushleft}
    \small 
    Each cell contains normalized values (energy, EDP)  when running the row-indexed network on an accelerator optimized for the column-indexed network.
 Color intensity
    indicates performance degradation severity: light yellow ($<$15\%), orange
    (15-50\%), and red ($>$50\%). Optimal accelerators were determined using
    our framework in Section~\ref{sec:methodology}, with homogeneous compute
    tiles selected for comparative clarity. Batch=1 and batch=4 are used for OPT-66B\_prefill. Framework variance enables accelerators optimized for one network to possibly perform better on others
\end{flushleft}
\end{table}

Furthermore, despite some accelerators incorporating heterogeneity, they operate at coarse granularity. Prefill-decode heterogeneity distinguishes only between phases~\cite{distservedisaggregatingprefilldecoding, splitwiseefficientgenerativellm}, while convolution-FC heterogeneity differentiates only between operation types~\cite{heterdata, BigLittle}. 
These coarse-grained approaches miss critical operator-level variations in computational patterns, memory access, and data reuse that exist within each phase or operation type.

This analysis necessitates accelerators with both network-level customization and operator-level heterogeneity—\glspl{nsic} that tailor their architectures to operator-specific memory requirements, batching characteristics, and utilization patterns.

Beyond these architectural insights, monolithic \glspl{nsic} face mounting economic challenges. 
The
\gls{nre} costs for custom silicon have risen dramatically with each new
process node~\cite{pioneering_amd, Chiplet_actuary}, with \SI{5}{\nano\meter}
designs now often exceeding \$100 million~\cite{UCIe_}. These escalating costs make
specialized accelerators economically viable only for the highest-volume
applications. 
\Gls{re} cost, implied by manufacturing yields, compounds this problem, as defects scale
superlinearly with IC area~\cite{how_small_is_too_small}, creating prohibitive
barriers to true architectural customization across diverse neural network
architectures.

Fortunately, chiplet-based systems present a promising solution: they enable network-level customization and operator-level heterogeneity through composable modular units~\cite{BigLittle, scar, Cambricon}, amortize \gls{nre} costs across multiple applications~\cite{Chiplet_actuary, AMD2024Chiplet, Monad, UCIe_}, and improve manufacturing yields through smaller die sizes~\cite{how_small_is_too_small, simba}.

However, determining which chiplets to include in the ecosystem and how to compose them into effective \glspl{nsic} remains nontrivial.
Suboptimal design decisions lead to two failure modes: excessive chiplet diversity that prevents adequate NRE amortization (too many unique chiplets with limited reuse opportunities), or insufficient chiplet coverage resulting in poor performance (missing critical chiplets or ineffective composition strategies). 
These challenges are fundamentally coupled—the chiplet pool's effectiveness depends on the quality of resulting \glspl{nsic}, while \gls{nsic} performance is constrained by available chiplets.
This circular dependency necessitates a chiplet ecosystem-accelerator codesign framework that simultaneously optimizes chiplet selection and \gls{nsic} composition.
While the maturing chiplet ecosystem—with standards like UCIe~\cite{UCIe_, UCIe} and universal interposers~\cite{lucie}—provides the infrastructure, systematic design methodologies for chiplet selection and composition remain underdeveloped.
\textbf{
To our knowledge, we are the first to address chiplet reuse through joint optimization of the chiplet ecosystem and accelerator design.
}

In this paper, we introduce Mozart, a comprehensive codesign framework that
systematically explores the chiplet-based accelerator design space to create
composite systems optimized for diverse AI deployment scenarios. Mozart
addresses operator-level architectural insights through three key techniques:
(1) chiplet-heterogeneity, which matches specialized chiplet types to different
computational patterns~\cite{scar}, (2) tensor fusion, which combines
operations to reduce data movement~\cite{deepfrack, tileflow, flat}, and (3)
tensor \& pipeline parallelism, which distributes computation across multiple
chiplets~\cite{HyPar}. The framework considers multiple optimization objectives
including energy efficiency, performance (\gls{edp}), and cost-effectiveness
(\gls{ec}, \gls{edpc}), enabling composite accelerators that excel in diverse deployment
contexts.

This paper makes several key contributions:
(1) \textbf{Mozart}, a chiplet ecosystem and accelerator co-design framework that breaks the circular dependency between chiplet pool composition and accelerator design;
(2) A comprehensive chiplet-based \gls{nsic} design methodology that translates operator-level architectural insights into concrete implementations, 
incorporating hierarchical algorithmic optimizations and integrated place-and-route validation to efficiently navigate the expansive chiplet design space;
(3) A constraint-aware optimization algorithm that generates tailored system-level solutions for diverse deployment contexts, 
from datacenter inference serving to autonomous vehicle perception, spanning \glspl{cnn}, \glspl{vit}, and language models.

\textbf{Upon publication, Mozart will be released as an open source design tool.}

\section{Operator Level Disaggregation}
\label{sec:motivation}
Modern neural network acceleration faces fundamental architectural challenges that motivate a shift from monolithic to chiplet-based designs. 
The disaggregated nature of chiplet-based designs motivates us to consider how operator-level disaggregation can address the growing inefficiencies in current accelerator architectures. 
We employ roofline models \cite{10.1145/1498765.1498785} for first-order analysis to demonstrate our architectural insights. 
In our section, the memory pool includes DDR5, LPDDR5, GDDR7, and HBM3E, covering mainstream memory modules. 
For computing chiplets, we consider a set of PE arrays ranging from 64×64 to 512×512.

\begin{insightbox}
\textbf{Insight 1: There is no memory wall, only compute-memory mismatches}
\end{insightbox}
The widely-cited ``memory wall'' \cite{10.1145/1498765.1498785}  in accelerator design assumes uniform memory requirements across all operations. This system-level perspective, however, masks the significant heterogeneity in memory demands across individual operators. Each computational operator exhibits different compute-to-memory ratios, creating operator-specific memory requirements rather than a homogeneous system-wide constraint.

\textit{Architectural Implication:} This insight suggests heterogeneous memory architectures tailored to operator-specific bandwidth requirements, enabling substantial system-level cost reductions without performance degradation. 

\begin{figure}[h]
    \centering
    \includegraphics[width=1.02\linewidth]{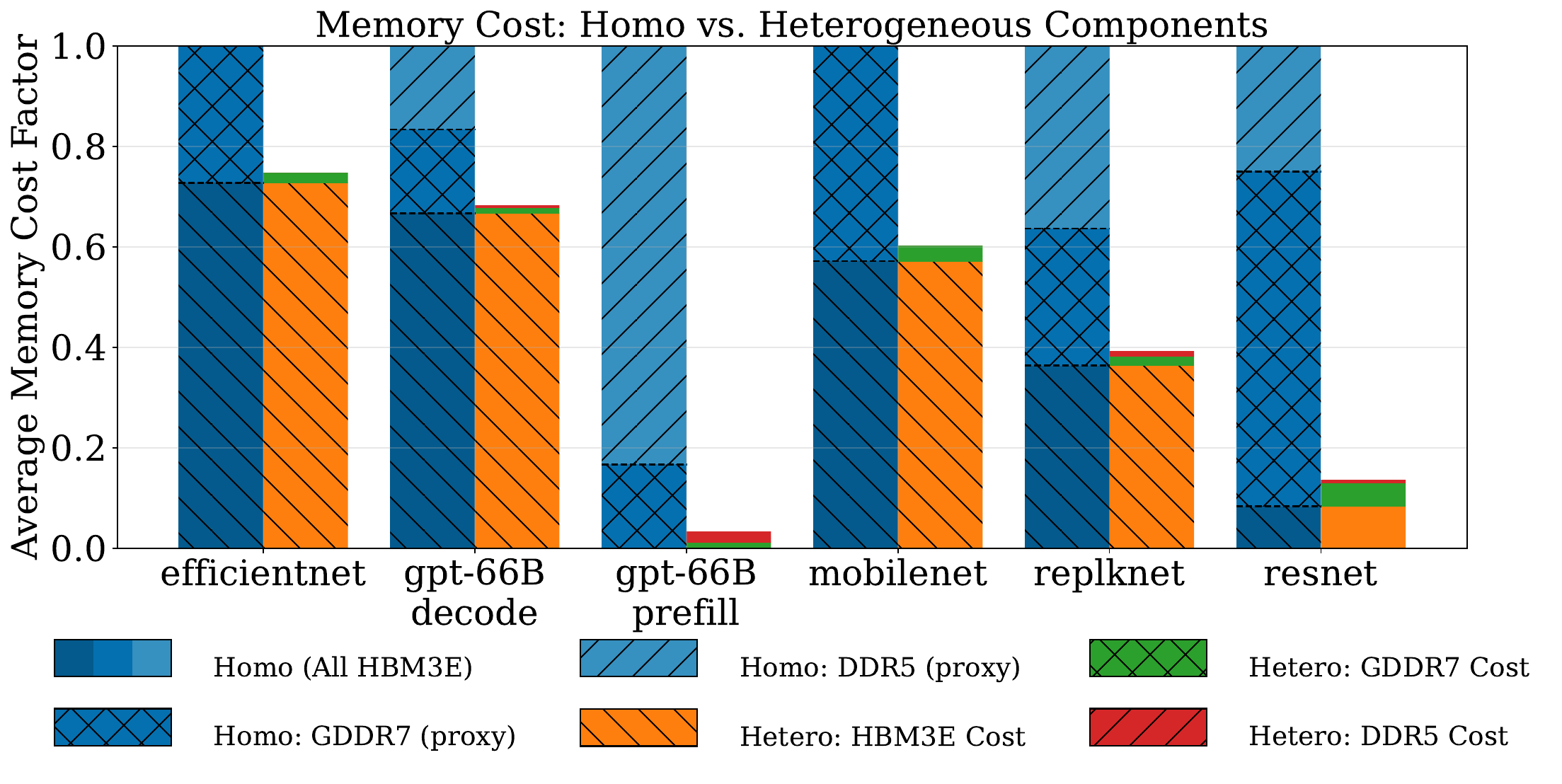}
    \caption{Heterogeneous memory systems enable significant cost optimization
    without performance degradation. Moving from homogeneous HBM3E to strategic
    combinations of HBM3E, GDDR7, and DDR5 maintains identical latency
    performance while achieving memory cost reductions of 25.4-96.7\% across
    CNN and GPT models through operator-specific memory allocation based on
    compute vs. memory-bound classifications.
    Memory costs are from~\cite{wikipedia_hbm}\cite{wikipedia_lpddr}\cite{samsung_k4z80325bc_datasheet}\cite{jedec_hbm3_2022}.
    }
    \label{fig:insight1_memory}
\end{figure}

As demonstrated in Figure~\ref{fig:insight1_memory}, moving from homogeneous HBM3E memory systems to heterogeneous memory architectures combining HBM3E, GDDR7, and DDR5 maintains identical latency performance across neural network models while achieving memory cost reductions of 25.4-96.7\% across CNNs and GPTs. Operators can be categorized as compute-bound or memory-bound, suggesting strategic memory allocation where compute-bound operators utilize cost-effective alternatives to expensive HBM3 without performance degradation.

\begin{insightbox}
\textbf{Insight 2: Universal batching sweet spot doesn't exist}
\end{insightbox}
Current system design assumes there exists an optimal batch size for neural network execution. While recent disaggregated prefill-decode architectures recognize phase-level differences, this assumption still ignores the fundamental heterogeneity in how different operators respond to batching within each phase. 
\Acrlongpl{bao} (e.g., attention operators) derive no benefit from batching since they cannot reuse weights across samples.
\Acrlongpl{bso} (e.g., projections) benefit from batching while memory-bound, but experience diminishing returns once they become compute-bound.

\textit{Architectural Implication:} This insight motivates fine-grained batch scheduling at the operator level. 

\begin{figure}[h]
    \centering
    \includegraphics[width=1.02\linewidth]{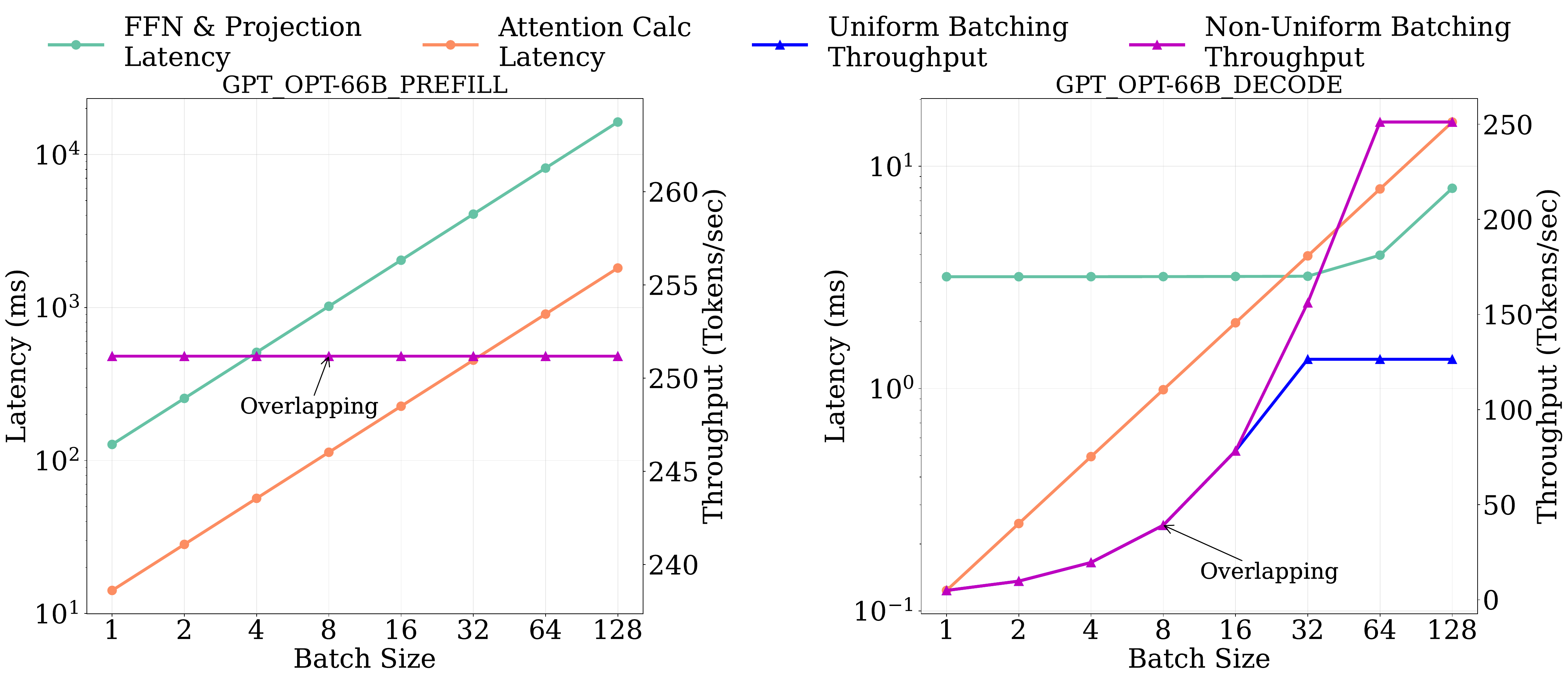}
    \caption{Batch scaling behavior varies dramatically across LLM operations,
    revealing operator-level heterogeneity that contradicts system-wide
    batching assumptions. Batching curves correspond to throughput scaling
    (right axis) while layer curves show latency scaling (left axis). 
    % Prefill operations scale linearly with batch size providing no throughput benefit,
    % while decode operations exhibit heterogeneous scaling where some layers
    % benefit from batching and others do not, motivating fine-grained batch
    % scheduling. Input text length: 2048 tokens.
    }
    \label{fig:insight2_batching}
\end{figure}

As demonstrated in Figure~\ref{fig:insight2_batching}, analysis of LLM workloads reveals these fundamental differences. LLM prefill operations scale linearly with batch size—execution latency doubles when batch size doubles while throughput remains constant—indicating no computational benefit from larger batches. In contrast, decode operations exhibit heterogeneous behavior: some operators scale linearly, indicating  no batch benefit, while others scale sublinearly, achieving increased throughput with larger batch sizes.

Current LLM serving systems (DistServe~\cite{distservedisaggregatingprefilldecoding}, SplitWise~\cite{splitwiseefficientgenerativellm}, WSC-LLM~\cite{10.1145/3695053.3731101})  apply uniform batching within each phase, capturing heterogeneity only at phase level rather than operator level. This wastes computational resources on batch-agnostic operations while underutilizing batch-sensitive ones.
We propose an operator-level heterogeneous batching strategy that employs small batch sizes with high tensor parallelism for batch-agnostic operators to mitigate the linear scaling of pipeline stage latency, while utilizing large batch sizes with low tensor parallelism for batch-sensitive operators to maximize weight reuse.

\begin{insightbox}
\textbf{Insight 3: The latency-goodput tradeoff is constrained by application requirements}
\end{insightbox}
System designers have long considered using batching to alleviate I/O bottlenecks and increase GPU goodput (utilization)~\cite{distservedisaggregatingprefilldecoding, sheng2023flexgenhighthroughputgenerativeinference, 280922}. 
However, this improvement comes at a cost: batching inevitably increases latency.
The latency-goodput tradeoff is fundamentally constrained by application requirements.
Some applications, such as autonomous vehicles, impose stringent latency requirements that limit batching opportunities.
In other cases, applications composed of multiple interdependent sub-services may enforce an even tighter latency constraint—for instance, the draft model in speculative decoding must decode significantly faster than the large model to enable timely batched verification.

\textit{Architectural Implication:} This insight reveals that the latency-goodput tradeoff may be infeasible for certain applications. 
In contrast, operator-level disaggregation enables latency-goodput decoupling—instead of relying on batching to increase goodput, high utilization can be achieved by replacing underutilized large chiplets with smaller, more efficient ones.
Interactive AI applications demonstrate this principle, as shown in Table~\ref{fig:insight3_latency}. While batching can improve goodput, it significantly increases \gls{ttft}, creating unacceptable delays for real-time applications that require immediate response. 
Operator-level disaggregation directly addresses the tension between goodput optimization and latency requirements in modern serving systems, enabling efficient processing for applications like interactive chatbots, real-time translation, and autonomous systems that require sub-second response times.

\begin{table}[h]
    \centering
    \begin{tabular}{|l|c|c|c|}
        \hline
        \textbf{Metric (GPT-66B)} & \textbf{No Batching} & \textbf{Batching} & \textbf{Hetero} \\
        \hline
        TTFT  & 3.295s {\cmark} & 26.362s {\cross} & 3.295s {\cmark} \\
        \hline
        Utilization & 23.8\% & 52.8\% & 88.6\% \\
        \hline
        Cost per tokens & 1 & 0.45 & 0.268 \\
        \hline
    \end{tabular}
    \vspace{1\baselineskip}
    \caption{\gls{ttft} analysis reveals the fundamental trade-off between goodput/batching and latency. Operator-level disaggregation enables heterogeneous architectures to achieve high goodput while maintaining low \gls{ttft}, critical for interactive AI applications.}
    \label{fig:insight3_latency}
    
\end{table}

\begin{insightbox}
\textbf{Insight 4: One size fits none: general-purpose accelerators excel at nothing}
\end{insightbox}
Accelerator designers pursue ``general-purpose'' architectures that can handle diverse neural network operators efficiently. This approach inherently creates architectural compromises that compound across different operation types. 
As
shown in Table~\ref{tab:intranet}, an accelerator optimized for convolutions with spatial data reuse performs poorly on attention mechanisms with different access patterns, while designs optimized for element-wise operations struggle with reduction operations requiring different PE array configurations and dataflows.

\textit{Architectural Implication:} This insight suggests operator-specific acceleration where each computational pattern receives dedicated optimization. 
Operator-level disaggregation suggests integration of heterogeneous accelerators, including varying dataflow patterns, processing element array sizes, and memory hierarchies, potentially delivering superior performance and energy efficiency compared to homogeneous designs forced to compromise across diverse operator requirements.

\begin{insightbox}
\textbf{Insight 5: Silicon real estate follows real estate rules: location (perimeter) beats size (area)}
\end{insightbox}
Accelerator scaling focuses on increasing total silicon area to improve performance, assuming larger chips deliver proportionally higher capability. This perspective ignores the geometric constraint that memory bandwidth scales with chip perimeter, not area. As chips grow larger, the perimeter-to-area ratio decreases, creating a fundamental scaling bottleneck that area alone cannot solve.

\textit{Architectural Implication:} This insight suggests disaggregation strategies that increase perimeter relative to area. 
A monolithic chip has limited perimeter available for memory interfaces, constraining total bandwidth regardless of internal compute density. By disaggregating the same total area into multiple smaller chiplets, the combined perimeter increases substantially --- enabling more memory interfaces and higher aggregate bandwidth. For example, disaggregating a single large square chip into $N$ smaller square chiplets increases the total perimeter by $\sqrt{N}\times$, potentially increasing the available memory bandwidth without additional silicon cost. 
This geometric advantage is particularly valuable for memory-bandwidth-limited AI workloads, enabling higher throughput per unit area and better scaling characteristics as model sizes continue to grow.

\begin{figure}[htbp]
    \centering
    \includegraphics[width=0.5\textwidth]{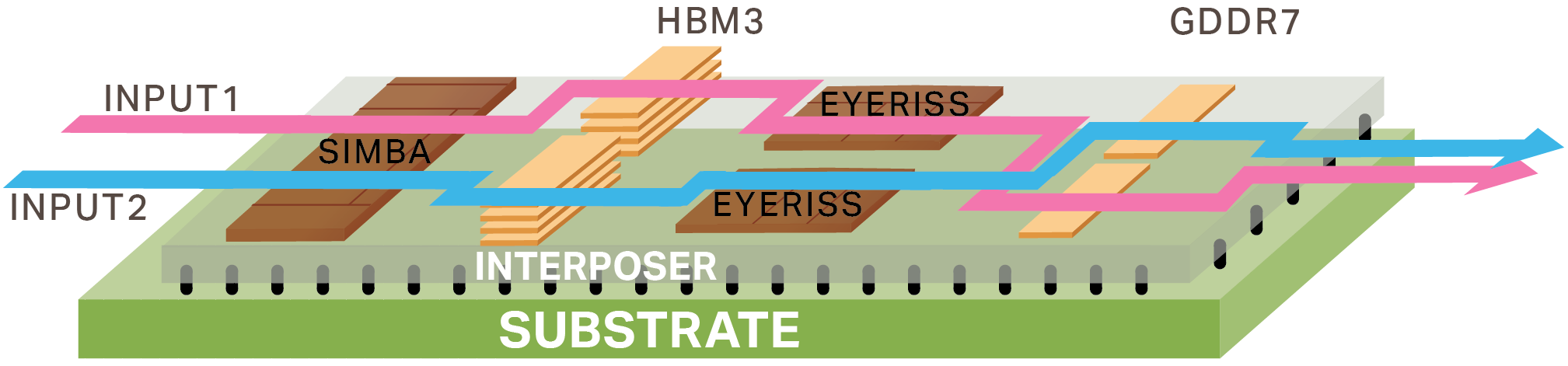}
    \caption{Architecture template of Mozart, showing \gls{db} for
    stall-free pipeline execution and token passing for memory access arbitration.
    \label{fig:arch}}
\end{figure}

These insights collectively motivate the systematic design of chiplet-based \glspl{nsic} and inform chiplet ecosystem development through their inherent coupling—effective \gls{nsic} design requires a well-curated chiplet ecosystem, while ecosystem composition depends on the requirements of target \glspl{nsic}.

By enabling operator-level disaggregation,
heterogeneous memory allocation, fine-grained batch scheduling, and geometric 
memory bandwidth scaling, our approach addresses the fundamental inefficiencies
in current monolithic accelerator designs while maintaining the performance
benefits of specialized hardware.

% \section{Architectural Diversity and Acceleration Challenges}
% \label{sec:background}
% \import{background}{background}

%\section{Existing Accelerator Design Frameworks}
\section{Survey of Existing Work}
\label{sec:problem_formulation}

Prior accelerator design efforts have focused primarily on dataflow mapping and
hardware co-optimization. Early frameworks like \cite{timeloop,cosa,maestro}
explored intra-layer mapping, while more advanced approaches such as
\cite{tileflow,deepfrack,fusemax} extended to layer fusion with analytical
optimization under fixed architectures. As shown in Table~\ref{other_works},
most existing frameworks support only a subset of the full design space and
typically assume homogeneous accelerator architectures.

While works like \cite{scar} and \cite{gemini} share some features with our
approach, significant differences remain in their implementations. To our
knowledge, no prior work simultaneously supports: \textbf{(1)} heterogeneous
chiplet selection, \textbf{(2)} mapping-fusion-parallelism co-optimization, and
\textbf{(3)} monetary cost modeling. Our framework is the first to combine
these dimensions, enabling end-to-end hardware-software codesign for emerging
workloads including Transformers and diffusion models, while explicitly
accounting for dollar cost.

\begin{table*}[t]
    \centering
    \caption{Existing Neural Network Accelerator Design Frameworks}
    \label{other_works}
    \scriptsize
    \resizebox{\linewidth}{!}{
    \begin{tabular}{|C{2.5cm}|C{1.3cm}|C{1.6cm}|C{1.3cm}|C{1.0cm}|C{1.5cm}|C{1.5cm}|C{1.8cm}|C{1.0cm}|C{1.9cm}|C{1.9cm}|C{1.9cm}|}
    \hline
        \multirow{2}{*}{Framework} & \multicolumn{2}{c|}{Hardware} & \multicolumn{4}{c|}{Software} & \multicolumn{5}{c|}{Chiplet} \\ \cline{2-12}
        ~ & Accelerator DSE & Heterogeneity & non-uniform batching & Tensor Fusion & Pipeline Parallelism & Tensor Parallelism & Chiplet Based System & Dollar Cost & Ecosystem Co-Design & Chiplet Floorplanning & Chiplet Ecosystem \\ \hline
        Timeloop\cite{timeloop} & \cross & \cross & \cross & \cross & \cross & \cross & \cross & \cross & \cross & \cross & \cross \\ \hline
        MAESTRO\cite{maestro} & \cmark & \cross & \cross & \cross & \cross & \cross & \cross & \cross & \cross & \cross & \cross \\ \hline
        CoSA\cite{cosa} & \cross & \cross & \cross & \cross & \cross & \cross & \cross & \cross & \cross & \cross & \cross \\ \hline
        Chimera\cite{chimera} & \cross & \cross & \cross & \cmark & \cross & \cross & \cross & \cross & \cross & \cross & \cross \\ \hline
        Tileflow\cite{tileflow} & \cross & \cross & \cross & \cmark & \cmark & \cross & \cross & \cross & \cross & \cross & \cross \\ \hline
        SET\cite{set} & \cross & \cross & \cross & \cmark & \cross & \cross & \cross & \cross & \cross & \cross & \cross \\ \hline
        DeFiNES\cite{defines} & \cmark & \cross & \cross & \cmark & \cross & \cross & \cross & \cross & \cross & \cross & \cross \\ \hline
        FLAT\cite{flat} & \cross & \cross & \cross & \cmark & \cross & \cross & \cross & \cross & \cross & \cross & \cross \\ \hline
        Fusemax\cite{fusemax} & \cross & \cross & \cross & \cmark & \cmark & \cross & \cross & \cross & \cross & \cross & \cross \\ \hline
        DeepFrack\cite{deepfrack} & \cross & \cross & \cross & \cmark & \cross & \cross & \cross & \cross & \cross & \cross & \cross \\ \hline
        SoMA\cite{soma} & \cmark & \cross & \cross & \cmark & \cross & \cross & \cross & \cross & \cross & \cross & \cross \\ \hline
        SCAR\cite{scar} & \cross & \cmark & \cross & \cross & \cmark & \cmark & \cmark & \cross & \cross & \cross & \cross \\ \hline
        Mind the Gap\cite{mind_the_gap} & \cmark & \cross & \cross & \cmark & \cross & \cross & \cross & \cross & \cross & \cross & \cross \\ \hline
        DoSA\cite{dosa} & \cmark & \cross & \cross & \cross & \cross & \cross & \cross & \cross & \cross & \cross & \cross \\ \hline
        Stellar\cite{stellar} & \cmark & \cross & \cross & \cross & \cmark & \cross & \cross & \cross & \cross & \cross & \cross \\ \hline
        LLMCompass\cite{llmcompass} & \cmark & \cross & \cross & \cross & \cmark & \cmark & \cross & \cmark & \cross & \cross & \cross \\ \hline
        Explainable\cite{explainable_dse} & \cmark & \cross & \cross & \cross & \cross & \cross & \cross & \cross & \cross & \cross & \cross \\ \hline
        DFModel\cite{dfmodel} & \cmark & \cross & \cross & \cmark & \cmark & \cmark & \cross & \cmark & \cross & \cross & \cross \\ \hline
        Cocco\cite{cocco} & \cmark & \cross & \cross & \cmark & \cross & \cross & \cross & \cross & \cross & \cross & \cross \\ \hline
        Gemini\cite{gemini} & \cmark & \cross & \cross & \cmark & \cmark & \cmark & \cmark & \cmark & \cross & \cross & \cross \\ \hline
        MOZART & \cmark & \cmark & \cmark & \cmark & \cmark & \cmark & \cmark & \cmark & \cmark & \cmark & \cmark \\ \hline
    \end{tabular}}
\end{table*}

\section{The Mozart Ecosystem-Accelerator Codesign Framework}
\label{sec:methodology}
We implement a deep pipeline architecture to showcase the operator-level heterogeneity (Figure~\ref{fig:arch}). 
Network layers are mapped to dedicated pipeline stages, with \gls{tp} increasing processing efficiency. 
Inter-stage
communication occurs through carefully-selected buffers, with costs modeled similarly
to~\cite{simba}.
\gls{db} techniques ensure continuous pipeline execution. Bus contention
is managed through token
passing arbitration.

\begin{figure}
    \includegraphics[width=\linewidth]{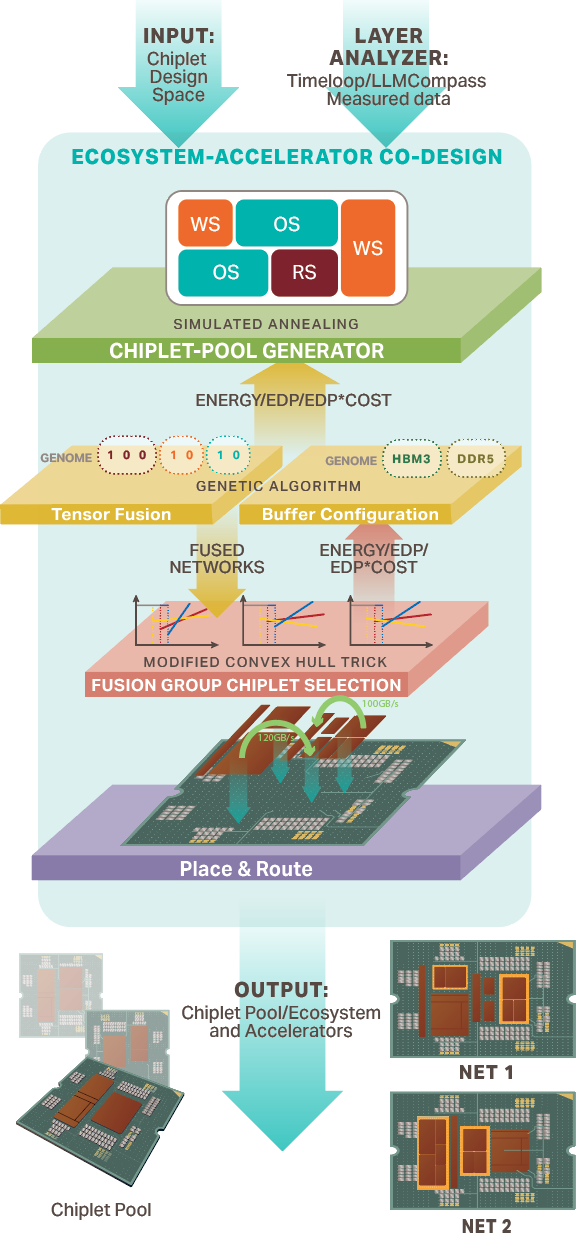}
    \caption{\label{fig:framework}
    Mozart's four-layer hierarchical framework: simulated annealing for chiplet pool composition, genetic algorithm for tensor fusion and buffer configuration, modified convex hull for chiplet selection, and place-and-route for physical implementation.
    }
\end{figure}

Mozart employs a hierarchical design space exploration framework to systematically compose chiplet-based accelerators (Figure~\ref{fig:framework}). The framework takes chiplet configurations, target neural networks, and optimization objectives (\gls{edp}, \gls{edpc}) as inputs, using Timeloop~\cite{timeloop,accelergy} and DeepFrack~\cite{deepfrack} for performance modeling.

The framework operates at four levels:
\textbf{Layer 1}: Simulated annealing explores chiplet pool compositions;
\textbf{Layer 2}: Genetic algorithm identifies tensor fusion strategies and buffer configurations;
\textbf{Layer 3}: Modified convex hull\cite{Convex1, Convex2} selects optimal hardware-software mappings;
\textbf{Layer 4}: Place-and-route determines physical implementation.

Performance metrics flow through the hierarchy to evaluate solutions at each level, ultimately providing effective chiplet ecosystem, optimized accelerators, and physical feasibility.

\subsection{Simulated Annealing for Chiplet Pool Composition}
\label{sec:sa}
We employ simulated annealing to explore effective chiplet pool compositions, with each pool evaluated on the lowest achievable performance metrics of accelerators constructed from it.

Each iteration generates candidate pools by modifying chiplet configurations: transitioning between dataflows (Row-Stationary, Weight-Stationary, Output-Stationary), adjusting PE array dimensions, and reconfiguring buffer capacities. Neighboring pools with similar architectures exhibit comparable performance characteristics, creating a well-formed optimization landscape suitable for simulated annealing.

% We select simulated annealing specifically because it naturally accommodates
% our definition of neighboring chiplet pools. Two chiplet pools are considered
% neighboring if they exhibit a high degree of similarity in their constituent
% chiplets. This similarity is characterized by the same dataflow pattern,
% comparable PE array dimensions, and global buffer capacities.

% Empirically speaking, neighboring chiplet pools—as defined by our architectural
% similarity metrics—tend to exhibit comparable performance characteristics. This
% property creates a well-formed optimization landscape with gradual transitions
% between performance regions, making it particularly suitable for the simulated
% annealing algorithm's exploration strategy.

\subsection{Evolutionary Search for \gls{tf} and Memory Allocation}
\label{sec:es}

We employ evolutionary search to simultaneously optimize tensor fusion grouping and pipeline buffer configurations, selecting appropriate memory types (HBM3, GDDR7, DDR5, and LPDDR5) that match the bandwidth requirements of each fusion group given the chiplet computing capacity.
To accelerate convergence, we leverage roofline models to seed the search with promising buffer configurations based on compute-memory ratios.

Our genetic representation preserves high-quality fusion groups through crossover operations while incorporating domain-specific knowledge to prune the search space.
For instance, Alwani et al.~\cite{first_fuse} demonstrated that fusing early layers in deep networks like VGGNet significantly improves energy efficiency—we directly encode such empirically-validated patterns into our initial population and mutation operators.

\subsection{Modified Convex Hull Trick for Layer Codesign}
\label{sec:cht}
% summary sentence
% With the chiplet pool optimized through simulated annealing and layer fusion
% strategies determined by evolutionary search, a modified version of convex hull
% trick takes over to identify the optimal chiplet allocation and corresponding
% software mapping under pre-defined objectives.
% 
% To facilitate rigorous analysis of computational complexity, we establish the
% following notation: $M$ denotes the number of distinct chiplet and software
% mapping configurations available at each pipeline stage, $P$ represents the
% total number of pipeline stages in the accelerator architecture, and $Q$
% signifies the cardinality of the set of possible discrete pipeline stage
% latencies.

Our modified convex hull trick identifies optimal chiplet allocation and software mapping for each tensor fusion group. We use notation: $M$ = configurations per pipeline stage, $P$ = total pipeline stages, $Q$ = possible discrete stage latencies.

\subsubsection{Energy as Piecewise Affine Function}
\label{sec:cht_em}
Total energy decomposes into dynamic and static components: $E = E_{\text{dynamic}} + E_{\text{static}}$. In pipelined accelerators, static energy presents challenges as chiplets completing early still consume leakage power while waiting for other stages, creating interdependencies where locally optimal selections may not yield globally optimal configurations.

We formulate the energy model as a piecewise affine function:
\begin{equation}
  E(T) = 
    \begin{cases} 
      E_{\text{dynamic}} + P_{\text{static}} \times T & \text{if } T \geq \tcompute{} \\
      \infty & \text{if } T < \tcompute{}
    \end{cases}
  \label{eq:final_energy}
\end{equation}

Where $T$ represents pipeline stage latency and $\tcompute{}$ denotes execution time for the tensor fusion group. Since static energy constitutes up to 30\% of total power~\cite{drowsy} and power gating has break-even points of \SI{1.5}{\milli\second}~\cite{static}, maintaining pipeline balance through careful chiplet selection is crucial.

\subsubsection{Na\"ive Approach}
\label{sec:naive}
A na\"ive approach would require exhaustive enumeration of all possible chiplets and mapping combinations across stages, resulting in computational complexity of $O(M^P)$. Such exponential complexity is intractable given the numerous tensor fusion strategies and chiplet pool compositions to be searched.

% \begin{algorithm}
% \caption{\textbf{Na\"ive search for optimal accelerator configuration}}
% \label{algo:naive}
% \textbf{Input:} $P$ distinct layer fusion groups; $M$ chiplet and software mapping options per group
% 
% \textbf{Output:} Optimal accelerator configuration and objective value 
% \begin{algorithmic}[1]
% \Function{NaiveSearch}{$stage, config, value$}
%     \State Init $best\_value \gets \infty$, $best\_config \gets \emptyset$
% 
%     \Function{Search}{$stage, config, value$}
%         \If{$stage = P$}
%             \If{$value < best\_value$}
%                 \State $best\_value \gets value$
%                 \State $best\_config \gets config$
%             \EndIf
%             \State \Return
%         \EndIf
% 
%         \For{$i = 1$ to $M$}
%             \State Select chiplet/mapping configuration $i$ for stage $stage$
%             \State Update $config$ with this selection
%             \State Compute new $value$ based on updated configuration
%             \State \Call{Search}{$stage + 1$, config, value}
%         \EndFor
%     \EndFunction
% 
%     \State \Call{Search}{$0$, $\emptyset$, $0$}
%     \State \Return $best\_config$, $best\_value$
% \EndFunction
% 
% \end{algorithmic}
% \end{algorithm}

\subsubsection{Iso-latency Approach}
\label{sec:iso}

The combinatorial explosion in our search space stems from the interdependence
of choices at each pipeline stage. We overcome this through iso-latency
analysis, decomposing the problem into two phases: (1) identifying sub-optimal
accelerator configurations at each discrete pipeline latency value; and (2)
determining the global optimum by comparing these configurations.

The key insight is that when pipeline stage latency is fixed, dependencies
between stages are eliminated. This allows independent optimization of each
stage for any given latency, transforming the problem from $O(M^P)$ complexity
to $O(M \times P \times Q)$.

As established in Section~\ref{sec:cht_em}, energy consumption at each stage is
modeled as a piecewise affine function of latency. Finding the optimal
configuration becomes a matter of evaluating all applicable affine functions at
that latency and selecting the one yielding minimal energy.

When extending to \gls{ec}, \gls{edp} or \gls{edpc}, we multiply energy consumption by
the corresponding latency and cost factor. Since our analysis maintains
iso-latency invariants, this
multiplication preserves solution optimality.

\subsubsection{Iso-latency Approach \& Modified Convex Hull trick}
\label{sec:iso_cht}
\begin{figure}[h]
    \centering
    \begin{subfigure}[b]{0.48\columnwidth}
        \centering
        \includegraphics[width=\textwidth]{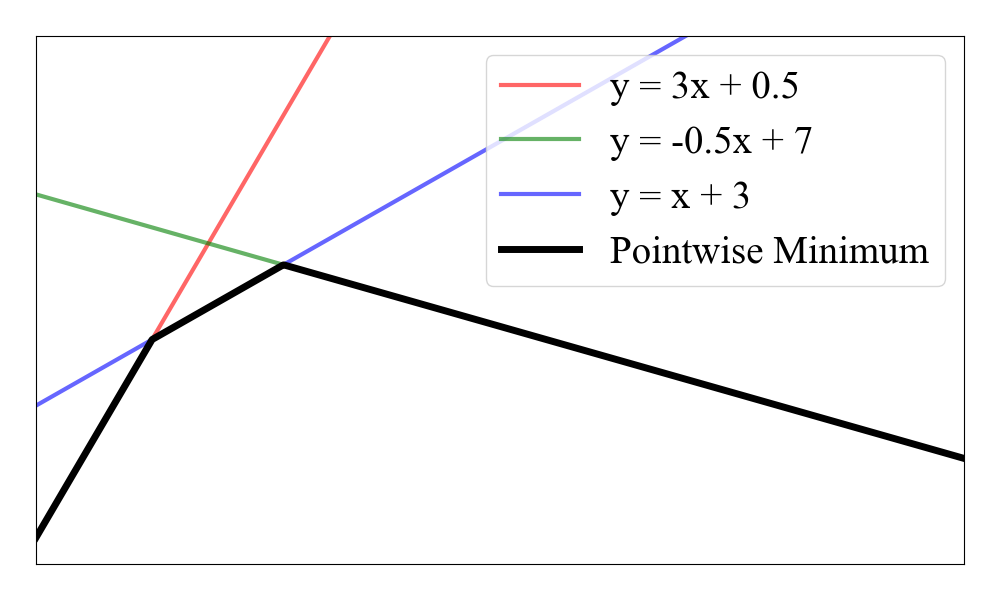}
        \caption{Affine functions \label{fig:cht} }
       
    \end{subfigure}
    \hfill
    \begin{subfigure}[b]{0.48\columnwidth}
        \centering
        \includegraphics[width=\textwidth]{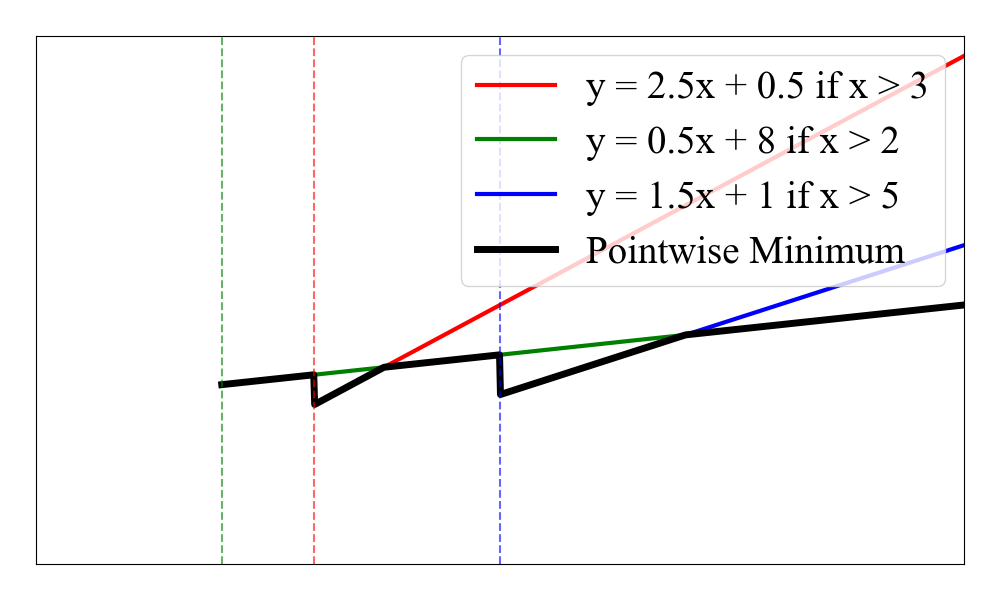}
        \caption{Piecewise affine functions \label{fig:mcht}}
        
    \end{subfigure}
    \caption{Convex hull trick for affine functions and piecewise affine functions \label{fig:cnt_show}}
    
\end{figure}

Although iso-latency analysis substantially reduces computational complexity,
further optimization is desirable given the extensive search space. 

The core challenge is finding the minimum value among piecewise affine functions at each pipeline stage latency. We employ the convex hull trick—a technique for efficiently determining which function attains minimal values (Figure~\ref{fig:cht}).

% First, every line on the hull provides the extremal value for some contiguous
% range of the independent variable, while lines not on the hull never provide
% the extremal value for any input and can be safely eliminated from
% consideration.
% 
% Second, all lines that form the hull have different slopes, with their slope
% values determining their relative positions on the hull. Since our objective is
% to find the minimal value, a line with lower slope appears on the hull to the
% left of one with a higher slope. This monotonic relationship enables efficient
% querying after preprocessing.

%The convex hull trick can be implemented in two distinct phases:
%\begin{enumerate} \item \textbf{Preprocessing phase:} Sort the affine functions
%            by slope and eliminate redundant functions that will never
%            contribute to the minimum value. This has time complexity $O(M \log
%            M)$.
%    
%\item \textbf{Query phase:} For each query point (pipeline stage latency),
%perform a binary search to determine the appropriate function based on slope,
%with the total time complexity $O(Q\log M)$. \end{enumerate}

Since we deal with piecewise affine functions (energy is infinite when latency is below $T_{cmp}$), we developed a modified convex hull trick (Figure~\ref{fig:mcht}) that maintains separate convex hulls for function subsets becoming active at different threshold points. 
Algorithm~\ref{algo:mcht} achieves $O(P \times (M \log M + Q \log M))$ complexity—a significant improvement for the parameter ranges relevant to our design space.

\begin{algorithm}
\caption{\textbf{Iso-latency with modified convex hull trick}}
\label{algo:mcht}
\footnotesize
\begin{description}
\item[\textbf{Input:}] Distinct \gls{tf} groups; chiplet and mapping options at group; discrete pipeline latency values.
\item[\textbf{Output:}] Optimal accelerator configuration and objective value
\end{description}
\begin{algorithmic}[1]
\Function{IsoLatencyWithConvexHullTrick}{}

    \State \texttt{bestVal} $\gets \infty$, \texttt{bestCfg} $\gets \emptyset$
    
    \For{\texttt{stage} $= 0$ \textbf{to} \texttt{P-1}}
        % \State \textit{// Preprocess affine functions for this pipeline stage}
        % \State Sort the $M$ affine functions by activation point (\texttt{T\_compute}) in ascending order
        % \State Initialize empty convex hull array \texttt{H[1...Q]}
        % \State \textit{// Build convex hulls for each activation point}
        \State \texttt{F\_sorted} $\gets$ \texttt{SortTCompute(\texttt{StageCfg})}
        \State \texttt{H[1…Q]} $\gets$ \texttt{InitEmptyHulls()}
        \For{\texttt{f\_i} in \texttt{F\_sorted}}
            \State \texttt{T\_i} $\gets$ \texttt{GetActivationPoint(f\_i)}
            
            \State \texttt{H[T\_i]} $\gets$ \texttt{H[T\_i$-1$]}
            
            \State \texttt{pos} $\gets$ \texttt{BinarySearchInsert(H[T\_i], f\_i)}
            
            \State \texttt{H[T\_i]} $\gets$ \texttt{RemoveIrrelevant(H[T\_i])}
            
            \State \texttt{InsertAt(H[T\_i], f\_i, pos)}
        \EndFor
        % \For{each function \texttt{f\_i} in sorted order}
        %     \State Let \texttt{T\_i} be the activation point of function \texttt{f\_i}
        %     \State Copy previous convex hull: \texttt{H[T\_i]} $\gets$ \texttt{H[T\_i-1]}
        %     \State Binary search to find insertion position for \texttt{f\_i} in \texttt{H[T\_i]}
        %     \State Remove irrelevant functions from \texttt{H[T\_i]}
        %     \State Add \texttt{f\_i} to \texttt{H[T\_i]} at the found position
        % \EndFor
    \EndFor
    
    \For{\texttt{T} in \texttt{pipeLatencys} }
        \State \texttt{curCfg} $\gets \emptyset$, \texttt{curVal} $\gets 0$
        \For{\texttt{stage} $= 0$ \textbf{to} $P-1$}
        
            % \State \textit{// Query optimal function at \texttt{T} for this stage}
            % \State Find appropriate convex hull \texttt{H[T']} where \texttt{T'} $\leq$ \texttt{T}
            % \State Binary search within \texttt{H[T']} to find minimal function at \texttt{T}
            \State \texttt{T'} $\gets$ \texttt{FindHull}(H, \texttt{T})
            \State \texttt{f\_min} $\gets$ \texttt{BinarySearchHull}(H[\texttt{T'}], \texttt{T})
            % \State \texttt{stage\_best\_value} $\gets$ value of minimal function at \texttt{T}
            % \State \texttt{stage\_best\_config} $\gets$ configuration corresponding to minimal function
            \State \texttt{StagebestVal}, \texttt{StagebestCfg} $\gets$ \texttt{Eval}(f\_min, \texttt{T})
            % \State Update \texttt{current\_config} with \texttt{stage\_best\_config}
            % \State \texttt{currValue} $\gets$ \texttt{currValue} + \texttt{stage\_best\_value}
            \State \texttt{curCfg} $\gets$  \texttt{curCfg} + \texttt{StagebestCfg}
            %\texttt{currConfig} $+$ \texttt{StagebestConfig}
            \State \texttt{curVal} $\gets$ \texttt{curVal} + \texttt{StagebestVal}
            %\texttt{currValue} $+$ \texttt{StagebestValue}
    \EndFor
        % \State Apply objective function modifier if necessary (for EDP or EDP $\times$ cost)
    %     \If{\texttt{Objective} $=$ \texttt{EDP}}
    %     \State \texttt{currValue} $\gets$ \texttt{currValue} $\times$ \texttt{T}
    % \EndIf
    % \If{\texttt{Objective} $=$ \texttt{EDP\_Cost}}
    %     \State \texttt{currValue} $\gets$ \texttt{currValue} $\times$ \texttt{C(curCfg)} 
    % \EndIf
        \State \texttt{curVal} $\gets$ \texttt{ObjFactor(curVal)}
        \If{\texttt{curVal} $<$\texttt{ bestVal}}
            \State \texttt{bestVal}, \texttt{bestCfg} $\gets$ \texttt{curVal}, \texttt{curCfg}
        \EndIf
    \EndFor
    \State \Return \texttt{bestCfg}, \texttt{bestVal}
\EndFunction
\end{algorithmic}
\end{algorithm}

The hierarchical framework coordinates co-optimization of chiplet composition, buffer configuration, tensor fusion, HW-SW mapping, and physical implementation. Performance metrics propagate bottom-up to guide optimization decisions while maintaining scalability across diverse objectives.

\subsection{Place and Route}
\label{sec:place_route}

The final layer of our hierarchical framework handles the physical implementation of chiplet-based accelerators through place and route. 
Given the chiplet allocation and interconnect requirements, this stage determines valid chiplet placement on the interposer and routes the inter-chiplet connections while satisfying physical design constraints.

The place and route layer focuses on constraint satisfaction, ensuring that: (1) all required chiplets fit within the interposer area, (2) inter-chiplet communication paths can be successfully routed, and (3) basic timing constraints are met. 
This step validates that the accelerator configurations identified by the upper optimization layers can be physically implemented. Subject to these feasibility constraints, the layer then minimizes interposer footprint to produce a more compact layout.

The place and route results provide feedback to the framework, confirming physical feasibility while updating latency and power estimates.
Thermal analysis and power delivery network validation remain as future work.

\subsection{Cost model}
\label{sec:cost_model}

We adopt the CATCH model~\cite{catch} to evaluate system cost under a unified RE and NRE framework. For RE cost, the model jointly considers wafer and lithography cost, yield, and packaging, models different packaging and interconnect technologies (e.g., hybrid bonding and TCB), and also accounts for memory controllers and PHYs.

Within this framework, the yield $Y_{\text{die}}$ decreases as area increases, leading to a superlinear rise in per-die cost:  
\[
C_{\text{die}} = \frac{K_{\text{die}}}{Y_{\text{die}}}.
\]  
Therefore, partitioning a large monolithic die into multiple smaller chiplets can significantly reduce manufacturing cost~\cite{how_small_is_too_small}.

In contrast, NRE cost is amortized over production volume and includes photomasks, validation hardware, and IP licensing, as well as the use of EDA tools and verification environments, and packaging/interposer design and prototyping. It also covers software-related investments, such as CPU--GPU software stack adaptation and optimization. These one-time costs must be incurred before mass production and have a significant impact on the overall cost structure. For a production volume $V$, the unit cost is:  
\[
C_{\text{unit}} = C_{\text{RE}} + \frac{C_{\text{NRE}}}{V}.
\]
Consequently, when the production volume is relatively small, the NRE cost becomes prohibitively high. Only under large-scale manufacturing does NRE cease to dominate the total cost.

\section{Evaluation Setup}
\label{sec:eva_setup}

We use TimeLoop v0.4~\cite{timeloop} and Accelergy~\cite{accelergy} for energy and performance simulation. 
Energy models for DRAM are calibrated using Cacti \cite{Catci_3dd, Catci_io}. 

We cover three canonical dataflow styles: output-stationary (OS), weight-stationary (WS), and row-stationary (RS). The architectural implementations follow those in \cite{eyeriss, simba, shidiannao}. 
Our workload suite spans CNNs (ResNet50, MobileNetV3, EfficientNet, ReplkNet-31) and transformers (\glspl{vit}, OPT-66B), with OPT-1.3B for speculative decoding evaluation. Representative regions are extracted for all benchmarks.

The experimental configuration are summarized in Table~\ref{tab:setup}.
\begin{table}[h]
\centering
\caption{Experimental Configuration}
\label{tab:setup}
\small
\begin{tabular}{@{}llll@{}}
\toprule
\multicolumn{2}{r}{\textbf{Chiplet Parameters}} & \multicolumn{2}{c}{\textbf{Algorithm Parameters}} \\
\midrule
Technology & 14\,nm & \multicolumn{2}{l}{\textbf{Simulated Annealing (SA)}} \\
Clock & 1\,GHz & Init. Temp & 1.0 \\
Tensor Par. & \{1, 2\} & Cooling Rate & 0.95 \\
GLB Scaling & \{1, 4, 9, 16\} & Iterations/Level & 5 \\
PE Scaling & \{1, 2, 3, 4\} & \multicolumn{2}{l}{\textbf{Genetic Algorithm (GA)}} \\
Dataflows & \{RS, OS, WS\} & Population & 10 \\
Bonding & \{2D, 2.5D\} & Generations & 10 \\
DRAM & \{LPDDR5, DDR5, & Mutation Rate & 0.2 \\
     & GDDR7, HBM3\} & Crossover Rate & 0.8 \\
Inter-Chip & \SI{1.3}{\pico\joule\per\bit} \cite{simba}& & \\
\bottomrule
\end{tabular}
\end{table}

\begin{paragraph}{GPU Baseline}
We compare our results against real GPU benchmarks obtained from an Nvidia A100 SXM4 40GB GPU. We implement all workloads and layers in PyTorch for GPU execution, and gather per-layer energy and latency with the NVML library.  To minimize kernel launch overheads and account for small kernels, we capture many kernel iterations as a CUDA graph to directly replay it on device, with pre-allocated buffers to avoid run-time memory allocation for fair comparison. 
We measure the total latency, as well as the energy consumption before and after the replay, and normalize for per-iteration results. Since the A100 GPU doesn't have an explicit manufacturer's suggested retail price, we set the cost of it to an optimistic estimate of \$10000, lower than any pricing we found from reliable retailers. In pipeline parallel execution, throughput is bottle-necked by the slowest layer in the pipeline - GPUs executing other layers will still draw power while waiting for the slowest layer despite being finished with their computation. To reflect actual power consumptions, we justify our measured powers assuming that all other GPU would draw idle power (measured at 45W for A100), until the slowest of them finishes.
\end{paragraph}

% \begin{itemize}
%     \item Experiment setup
%     \begin{itemize}
%         \item Introduction of models
%         \item Introduction of baseline
%     \end{itemize}
%     \item How many chiplets should be used with the objective of
%     \begin{itemize}
%         \item energy
%         \item \gls{edp}
%         \item \gls{edp}*cost
%         %\item sum(energy+\gls{edp}+\gls{edp}*cost)
%     \end{itemize}
%     \item Comparison between our solution and baselines
% \end{itemize}

\section{Evaluation}
\subsection{Codesign Framework}
\label{sec:eva_codesign}
Mozart's effectiveness stems from its ability to translate the five operator-level insights from Section~\ref{sec:motivation} into practical architectural benefits. We evaluate the framework across multiple dimensions to demonstrate how these insights manifest in real systems.

\begin{figure}[t]
    \centering
    \includegraphics[width=0.8\columnwidth]{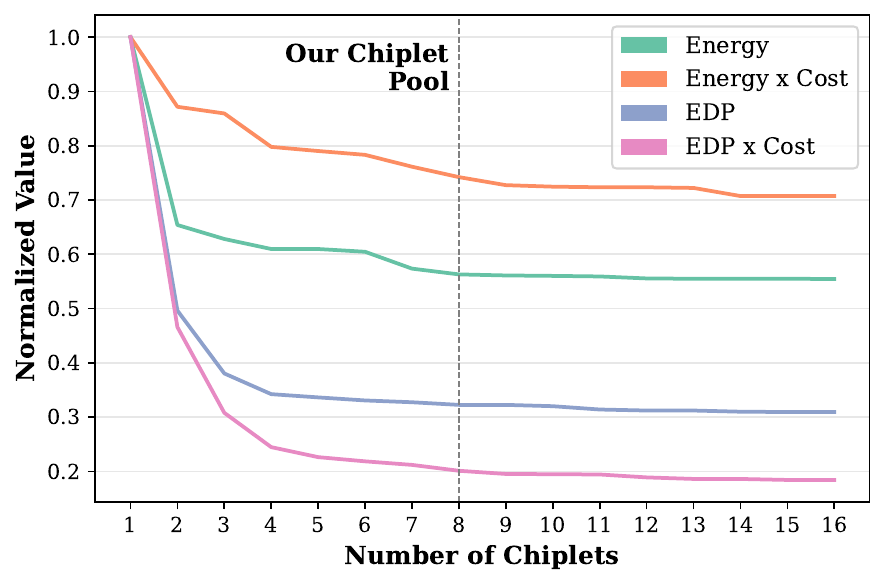}
    \caption{\label{fig:pool_size} Mozart constructs chiplet pools of varying sizes, optimized for different performance metrics. We identify 8 chiplets as the sweet spot, balancing performance gains with economic constraints—larger pools exhibit diminishing returns while increasing NRE costs.}
\end{figure}

\paragraph{Chiplet Pool Scaling and Composition}
We first examine how chiplet pool diversity affects system performance. 
As the number of specialized chiplet types increases, we measure improvements in energy, \gls{ec}, \gls{edp}, and \gls{edpc} across diverse networks. Figure~\ref{fig:pool_size} shows results normalized to a homogeneous architecture (chiplet pool with only 1 chiplet). 
While we see significant improvements when moving towards specialized architectures initially, adding more and more chiplets yields diminishing returns. Through this experiment, we identified a sweet spot of 8 chiplets in the tradeoff between performance and NRE costs. We use that as the optimal pool size that balances performance gains with economic constraints.

\paragraph{Architectural Comparison Study}
To demonstrate Mozart's advantages over existing approaches, we compare five architectural paradigms: \textbf{(1) GPU baseline} for general-purpose acceleration, \textbf{(2) Homogeneous ASIC (all networks)} using a single design for all workloads, \textbf{(3) Homogeneous \gls{nsic}} with network-specific homogeneous designs, \textbf{(4) Heterogeneous \gls{nsic} (chiplet pool)} implementing Mozart with operator-level heterogeneity, and \textbf{(5) Heterogeneous \gls{nsic} (unconstrained)} representing the theoretical upper bound with unlimited chiplet variety. 

We evaluate their performance on different workloads with four metrics: energy, \gls{edp}, \gls{ec}, and \gls{edpc} metrics. 
The latter two metrics aim to reflect \gls{tco} of different architectural paradigms.

\begin{figure*}[!t]
  \centering
  \includegraphics[width=\textwidth]{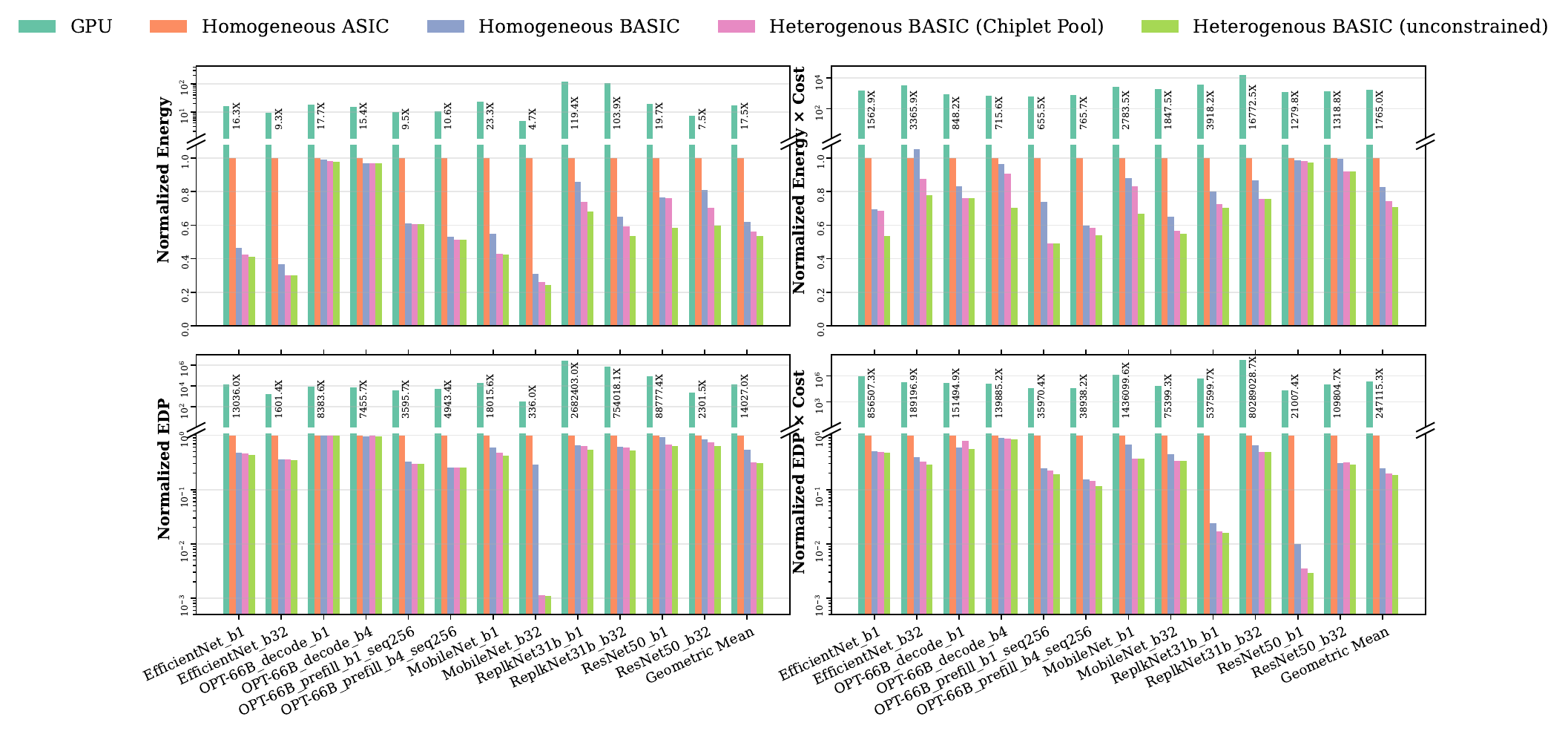} % or .png
  \caption{\label{fig:big_eval} Energy, \gls{ec}, \gls{edp}, and \gls{edpc} results of different architectural paradigms across different neural workloads, normalized to Homogeneous ASIC (all networks). }
\end{figure*}

Figure~\ref{fig:big_eval} presents the side-by-side results. 
GPU benchmarks are plotted on a broken axis to avoid distorting the scale of other paradigms. 
All paradigms demonstrate tremendous energy and EDP reductions relative to GPU, with homogeneous ASIC achieving 17.5× geometric mean energy savings and over 14,000× EDP savings. 
We attribute this improvement to the high utilization and lower overhead of ASICs compared to general-purpose compute.
One outlier is ReplkNet31b, where we used the naive PyTorch implementation without manual optimization. 
Due to the unique large convolutional kernels (31×31), the naive implementation performs poorly on GPUs. 
Excluding ReplkNet31b, homogeneous ASIC outperforms GPU with 12× geometric mean energy savings and over 5,568× \gls{edp} savings.

The transition from homogeneous ASIC to heterogeneous \gls{nsic} also yields significant savings in energy and reductions in \gls{edp} across all but decode workloads. 
Against unconstrained chiplet varieties, our chiplet pool of just 8 chiplets scores within 5\% for energy, \gls{edp}, and \gls{edpc}, and within 9\% in \gls{ec}.
Decode benefits less from specialized designs in terms of energy and EDP, but exhibits lower \gls{ec} and \gls{edpc}, which represents sizable economic savings.

It is worth noting, however, that costs in this metric only accounts for manufacturing costs, and the added NRE cost for designing and developing numerous chiplets renders the ideal paradigm impractical. 
In contrast, our 8 chiplet pool strategy offers a balance between NRE, operating costs, and performance. The results demonstrate how our systematic approach to chiplet-based acceleration resolves the dilemma between heterogeneity and reusability.

\begin{figure}[htbp]
    \centering
    \includegraphics[width=0.5\textwidth]{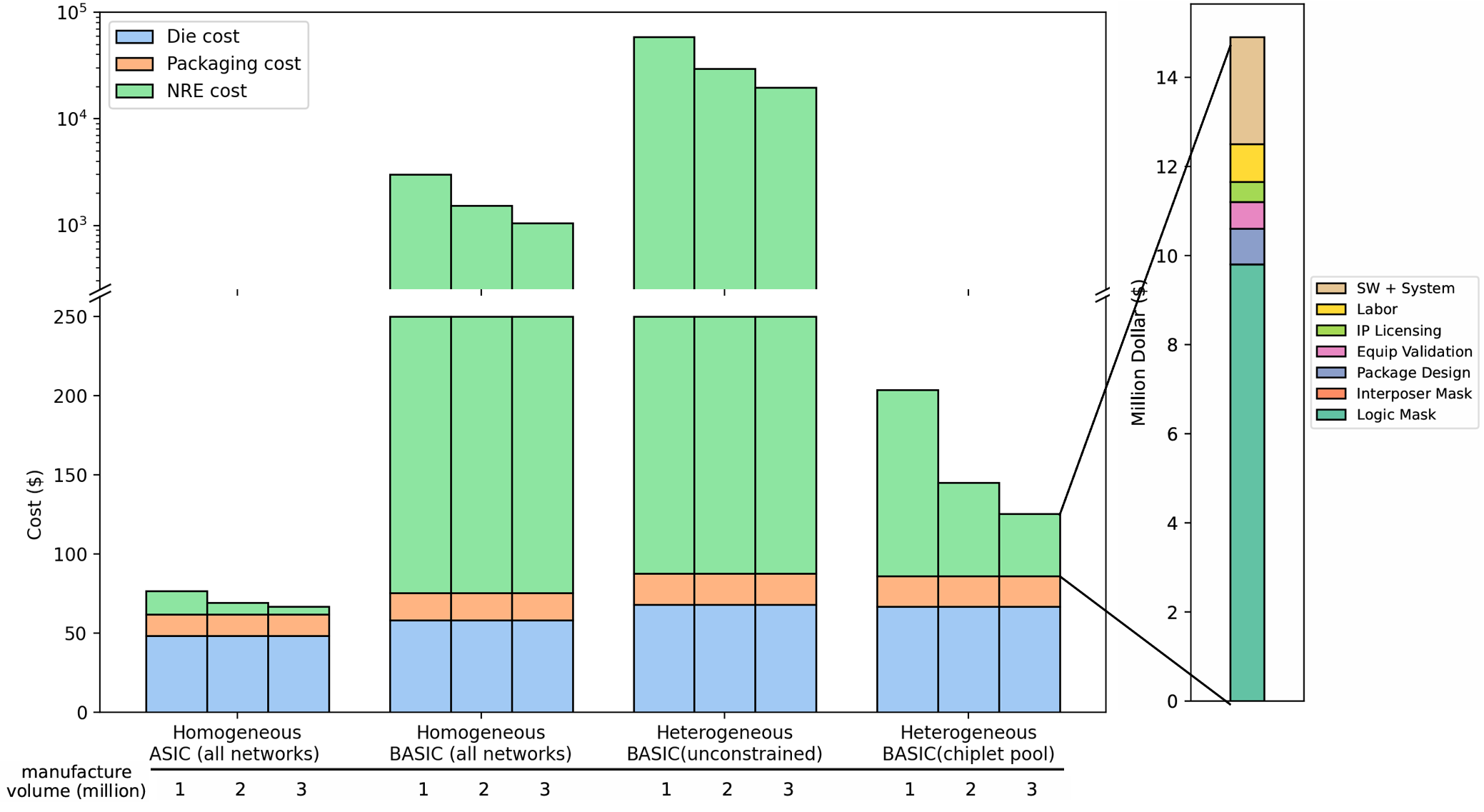}
    \caption{System cost breakdown under different manufacturing volumes and integration strategies (illustrated using ReplkNet31B), assuming a total of 200 different networks. The panel on the right further details the major components of NRE cost.}
    \label{fig:nre_cost}
\end{figure}

Figure~\ref{fig:nre_cost} presents the system cost structure of chiplet-based integration across different manufacturing scenarios. Within each strategy, the three adjacent bars represent manufacturing volumes of 1M, 2M, and 3M units, respectively. 
As shown, die and packaging costs remain relatively stable across strategies and scales, while NRE cost dominates the overall expenditure, especially at smaller scales. In contrast, the chiplet pool strategy achieves substantial cost reduction while maintaining effectiveness, highlighting its economic advantage and practical value in large-scale manufacturing.

% \paragraph{Framework Transferability}
% We evaluate how well Mozart-optimized chiplet pools generalize across different network architectures by training on a subset of models and testing on unseen networks. This study validates Insight 5's claim about geometric memory bandwidth advantages and demonstrates the reusability of specialized chiplets across diverse workloads.

% \paragraph{Ablation Study}
% We isolate the contributions of Mozart's three key techniques—tensor parallelism (TP), tensor fusion (TF), and heterogeneous memory systems—to quantify their individual and combined impact on system performance. This analysis directly validates the architectural insights from Section~\ref{sec:motivation} and demonstrates their practical implementation benefits.

\subsection{Case Study}
\label{sec:eva_system}

The following case studies demonstrate how Mozart's operator-level insights translate into practical benefits across diverse AI deployment scenarios that span from datacenter inference serving to energy-constrained and latency critical edge computing.
The latency requirement \cite{distservedisaggregatingprefilldecoding} for each case study is shown in Table~\ref{tab:lat}. 
We impose latency requirements to constraint the search process within our framework.

\begin{table}[htp]
\centering
\caption{Latency requirements of workloads}
\label{tab:lat}
\footnotesize
\begin{tabular}{@{}lccc@{}}
\toprule
\textbf{Application} & \textbf{TTFT (s)} & \textbf{TPOT (s)} & \textbf{E2E latency (s)} \\
\midrule
Chatbot OPT-66B & 2.5 & 0.15 & --- \\
Summarization OPT-66B & 15 & 0.15 & --- \\
Autonomous Vehicles ViT/CNN & --- & --- & 0.01--0.033 \\
\bottomrule
\end{tabular}
\end{table}

\subsubsection{Datacenter Large Language Model Serving}
\label{eva:sys_pd}

This case study demonstrates how Mozart addresses the fundamental challenges of datacenter LLM serving—reducing total cost of ownership (TCO) through improved energy efficiency and lower system costs, while meeting stringent quality of service (QoS) requirements for time-to-first-token (TTFT) and time-per-output-token (TPOT).

\paragraph{Standard LLM Serving}
State-of-the-art systems such as DistServe~\cite{distservedisaggregatingprefilldecoding} and SplitWise~\cite{splitwiseefficientgenerativellm} employ heterogeneous GPU configurations (e.g., A100, H100) and differentiated batching strategies for prefill and decode phases to accommodate their distinct computational characteristics. 
However, these systems maintain uniform computing resources and batching strategies within each phase.
This case study compares two approaches: (1) \textbf{DistServe}, which utilizes phase-level heterogeneous chiplets from the chiplet pool with uniform batching strategies, (2) \textbf{DistServe + Mozart}, which employs operator-level heterogeneous chiplets from a chiplet pool with non-uniform batching strategies.

\begin{figure}[t]
    \centering
    \includegraphics[width=1\columnwidth]{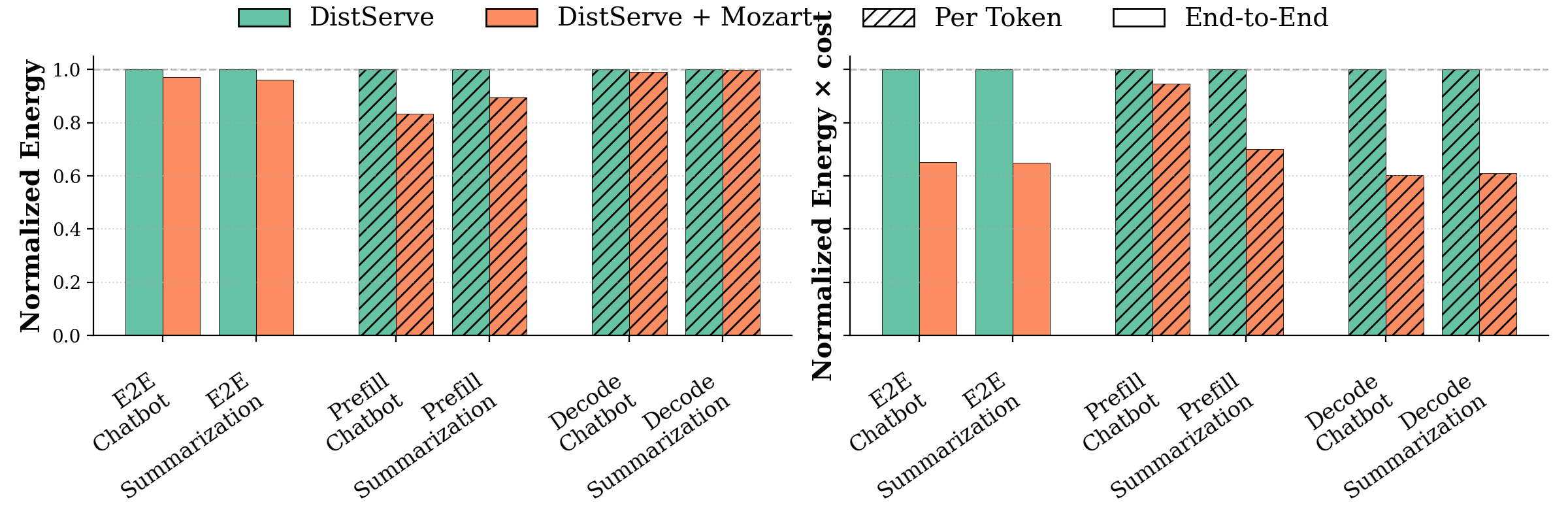}
    \caption{\label{fig:summ_and_chat} LLM end-to-end and per-token energy consumption. The framework explores various batching strategies without violating latency constraints. Both E2E per-request energy ($\times$ cost) and per-token energy ($\times$ cost) metrics are evaluated across different chiplet configurations.
}
\end{figure}

As shown in Figure~\ref{fig:summ_and_chat}, employing operator-level heterogeneous chiplets with non-uniform batching strategies yields a 15\% to 19\% reduction in energy consumption for the prefill stage. In terms of \gls{ec}, a 35\% to 39\% reduction is achieved for E2E requests. These improvements stem from two key factors: the increased batch sizes enabled by non-uniform batching strategies, and the strategic deployment of lower-cost memory and compute tiles for non-critical operators through operator-level heterogeneity.

\paragraph{Speculative Decoding Integration}
Mozart’s operator-level approach naturally extends to speculative decoding (SD), where a small
\emph{draft} model accelerates a large \emph{target} by proposing $k$ tokens per iteration for
batched verification~\cite{sd}. This setting makes the latency–throughput trade-off explicitly
\emph{operator-dependent}: the draft path is latency-critical, while the verifier path is
throughput-oriented. 
Following prior work~\cite{decoding_sd}, we evaluate OPT-66B (target) with
OPT-1.3B (draft), set \gls{tar} to 5.6 (with $k\!\ge\!5$), and cap realized speedup at
$2\times$ over non-SD by limiting the draft’s decode rate (thereby constraining draft latency).
We compare Mozart’s heterogeneous chiplet pool against a homogeneous chiplet baseline. Mozart
allocates latency-sensitive draft operators to speed-optimized chiplets and routes verifier
operators to throughput-optimized designs. We report throughput (speedup) and energy (including
\gls{ec}).

\begin{figure}[t]
  \centering
  \includegraphics[width=\linewidth]{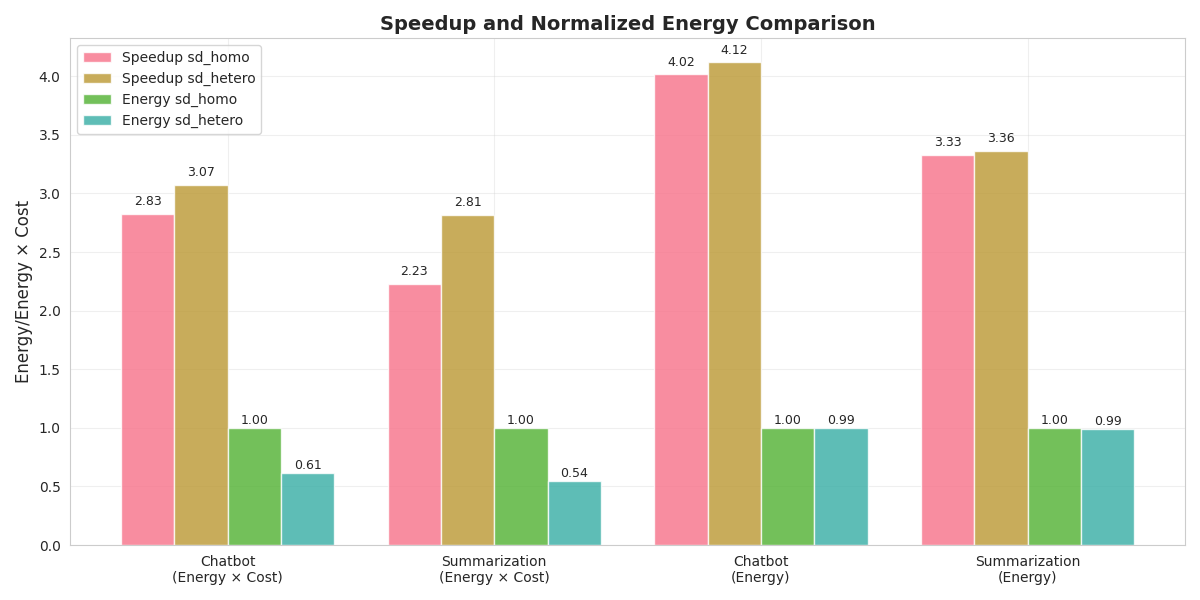}
  \caption{Speculative decoding results under a $2\times$ cap and $\mathrm{TAR}{=}5.6$:
  Mozart’s heterogeneous chiplet pool vs.\ a homogeneous chiplet baseline. We report
  throughput (speedup) and energy (including \gls{ec}) for \emph{Chatbot} and
  \emph{Summarization} in both cost-aware and performance-only settings; all runs meet
  TTFT/TPOT constraints.}
  \label{fig:sd-results}
\end{figure}

As seen in Figure~\ref{fig:sd-results}, under the same $2\times$ cap and $\mathrm{TAR}{=}5.6$, Mozart consistently outperforms the
homogeneous baseline. In cost-aware configurations, it increases throughput by \textbf{24.6\%}
on \emph{Chatbot} and \textbf{58.6\%} on \emph{Summarization}, while reducing energy by
\textbf{38.6\%} and \textbf{45.6\%}, respectively. In performance-only configurations, it
delivers smaller throughput gains
with near energy parity, and all settings satisfy the
TTFT/TPOT constraints.

% \hl{We will show that Mozart outperforms the existing GPU-based solution in terms of energy and cost efficiency while satisfying the stringent TTFT and TPOT requirements under different levels of token acceptance rate. }

\subsubsection{Edge Computer Vision for Autonomous Vehicles}
\label{eva:sys_vt}
This case study validates Mozart’s effectiveness under \emph{both} energy- and latency-constrained
edge scenarios, with a special focus on vehicle perception where computational efficiency directly
affects operational range. We evaluate under realistic autonomous-vehicle constraints—\emph{low
batch sizes} (typically 1 for real-time inference) and \emph{strict energy budgets}—highlighting
the practical relevance of Insights~1 and~4 from Section~\ref{sec:motivation}. We report
energy/Frame, real-time constraint satisfaction, and \gls{ec}/Frame to capture the energy–cost
trade space for in-vehicle deployment.

\textbf{Latency envelope.}
Autonomous driving perception-plan-ning stacks typically update at 10–12\,Hz,
implying an end-to-end (E2E) budget of roughly 80–100\,ms per cycle; this cadence is
commonly assumed in in-vehicle systems/architecture analyses \cite{Constraints}.
Within this envelope, detection is the time-critical stage. We adopt the community’s
30\,FPS ($\approx$33\,ms) “real-time” threshold \cite{yolo,efficientdet}, and note that
mobile/edge detectors can reach $\approx$10–12\,ms \cite{mobiledets}. Accordingly,
we evaluate two DET deadlines, $\tau_{\text{DET}}\in\{33\text{ ms},\,10\text{ ms}\}$.
\begin{figure}[t]
  \centering
  \includegraphics[width=\linewidth]{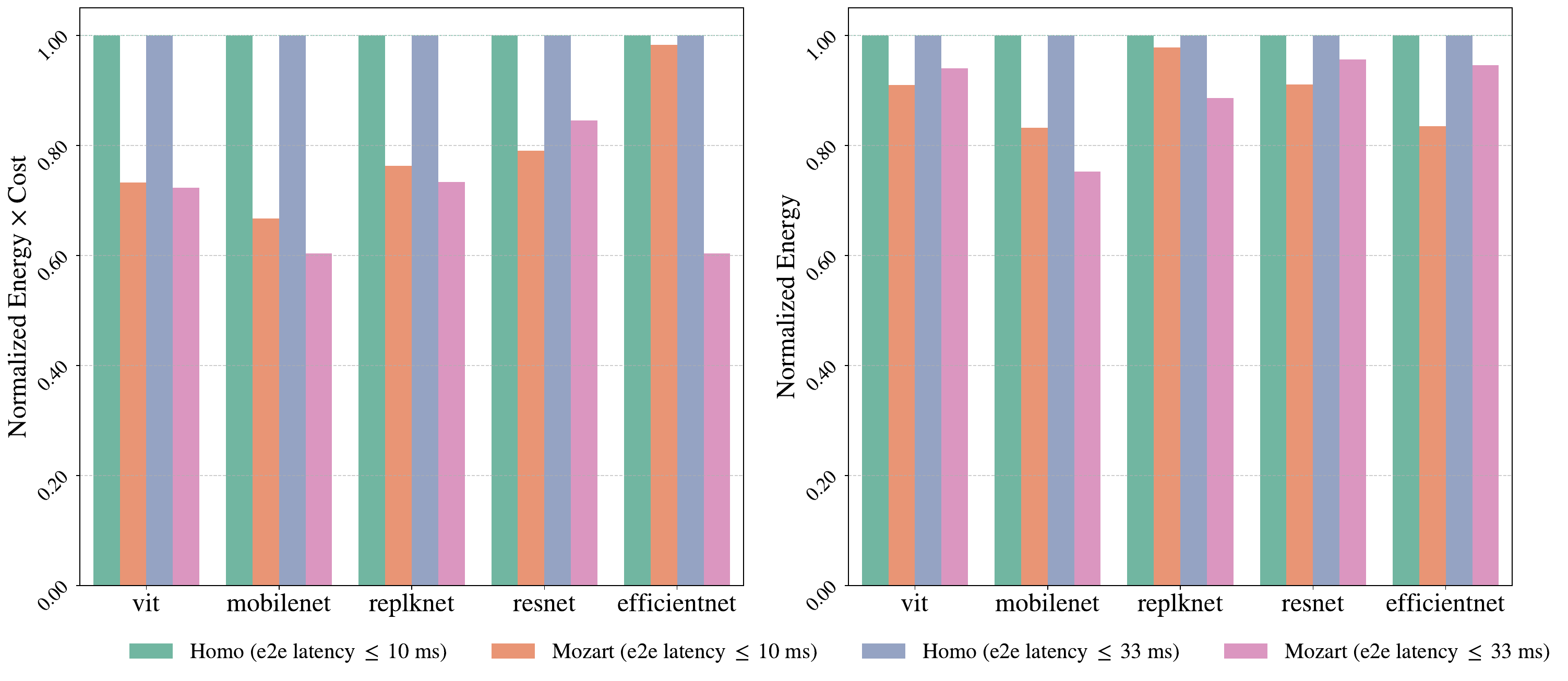}
  \caption{Normalized \gls{ec} (left) and normalized energy (right)
  under \emph{DET} deadlines of 10\,ms and 33\,ms. Bars are normalized to the
  homogeneous chiplet baseline. Mozart consistently reduces energy and
  \gls{ec} across \gls{vit}, MobileNet, RepLKNet, ResNet, and EfficientNet.}
  \label{fig:edge-det}
\end{figure}

Across backbones and under both DET deadlines (10\,ms and 33\,ms), Mozart lowers
\gls{ec} by \textbf{25.54\%} on average and reduces per-frame energy by
\textbf{10.53\%}, while meeting the E2E budget. The improvements are mainly due to
targeted heterogeneity and bandwidth–aware placement—consistent with
Insights~1 and~4 (Sec.~\ref{sec:motivation})—rather than aggressive frequency
scaling. The trend is stable across CNN/\gls{vit} and under the
typical low-batch settings vision workloads, indicating that operator-level
mapping avoids the cross-operator compromises inherent to homogeneous designs.
Under the tighter 10\,ms deadline, resources shift toward latency-critical stages
(as expected), but the relative energy and \gls{ec} advantages persist,
suggesting the approach remains effective even with reduced timing headroom.

\section{Conclusion}
\label{sec:conclusion}
This paper introduced Mozart, a chiplet ecosystem and accelerator codesign framework that addresses neural network acceleration by operating at the granularity of individual operators rather than entire networks. Our operator-level analysis revealed five critical insights that challenge conventional assumptions about memory requirements, batching effectiveness, and latency-goodput tradeoffs, demonstrating that these challenges manifest differently across individual computational patterns.

Through chiplet-heterogeneity, tensor fusion, and tensor parallelism, Mozart achieves 43.5\%, 25.4\%, 67.7\%, and 78.8\%
  savings in energy, \gls{ec}, \gls{edp}, and \gls{edpc} compared to traditional homogeneous accelerators while maintaining performance within 91\% to 95\% of monolithic designs. Crucially, just 8 strategically selected chiplet types can achieve these benefits, demonstrating economic viability through component reuse.

Case studies across datacenter LLM serving, and autonomous vehicle perception validate Mozart's effectiveness across contemporary AI architectures. For datacenter LLM serving, Mozart achieves 15-19\% energy reduction and 35-39\% energy-cost improvement through operator-level heterogeneity and non-uniform batching. In speculative decoding scenarios, Mozart achieves throughput improvements of 24.6\% for chatbot workloads and 58.6\% for summarization tasks, while reducing energy consumption by 38.6\% and 45.6\% respectively. For autonomous vehicle perception, Mozart reduces energy×cost by 25.54\% and energy by 10.53\% while meeting real-time constraints. The framework enables new deployment scenarios where specialized performance was previously economically unattainable, opening research directions in application-aware accelerator design that balance performance, energy efficiency, and economic considerations.

%%%%%%% -- PAPER CONTENT ENDS -- %%%%%%%%
%%%%%%%%% -- BIB STYLE AND FILE -- %%%%%%%%
\bibliographystyle{ACM-Reference-Format}
\bibliography{refs}

%%% -*-BibTeX-*-
%%% Do NOT edit. File created by BibTeX with style
%%% ACM-Reference-Format-Journals [18-Jan-2012].

\begin{thebibliography}{73}

%%% ====================================================================
%%% NOTE TO THE USER: you can override these defaults by providing
%%% customized versions of any of these macros before the \bibliography
%%% command.  Each of them MUST provide its own final punctuation,
%%% except for \shownote{}, \showDOI{}, and \showURL{}.  The latter two
%%% do not use final punctuation, in order to avoid confusing it with
%%% the Web address.
%%%
%%% To suppress output of a particular field, define its macro to expand
%%% to an empty string, or better, \unskip, like this:
%%%
%%% \newcommand{\showDOI}[1]{\unskip}   % LaTeX syntax
%%%
%%% \def \showDOI #1{\unskip}           % plain TeX syntax
%%%
%%% ====================================================================

\ifx \showCODEN    \undefined \def \showCODEN     #1{\unskip}     \fi
\ifx \showDOI      \undefined \def \showDOI       #1{#1}\fi
\ifx \showISBNx    \undefined \def \showISBNx     #1{\unskip}     \fi
\ifx \showISBNxiii \undefined \def \showISBNxiii  #1{\unskip}     \fi
\ifx \showISSN     \undefined \def \showISSN      #1{\unskip}     \fi
\ifx \showLCCN     \undefined \def \showLCCN      #1{\unskip}     \fi
\ifx \shownote     \undefined \def \shownote      #1{#1}          \fi
\ifx \showarticletitle \undefined \def \showarticletitle #1{#1}   \fi
\ifx \showURL      \undefined \def \showURL       {\relax}        \fi
% The following commands are used for tagged output and should be
% invisible to TeX
\providecommand\bibfield[2]{#2}
\providecommand\bibinfo[2]{#2}
\providecommand\natexlab[1]{#1}
\providecommand\showeprint[2][]{arXiv:#2}

\bibitem[{Advanced Micro Devices (AMD)}(2024)]%
        {AMD2024Chiplet}
\bibfield{author}{\bibinfo{person}{{Advanced Micro Devices (AMD)}}.} \bibinfo{year}{2024}\natexlab{}.
\newblock \bibinfo{booktitle}{\emph{AMD Chiplet Ecosystem}}.
\newblock \bibinfo{type}{White Paper}. \bibinfo{institution}{Advanced Micro Devices (AMD)}.
\newblock
\urldef\tempurl%
\url{https://www.amd.com/content/dam/amd/en/documents/solutions/technologies/chiplet-architecture-white-paper.pdf}
\showURL{%
\tempurl}


\bibitem[Alwani et~al\mbox{.}(2016)]%
        {first_fuse}
\bibfield{author}{\bibinfo{person}{Manoj Alwani}, \bibinfo{person}{Han Chen}, \bibinfo{person}{Michael Ferdman}, {and} \bibinfo{person}{Peter Milder}.} \bibinfo{year}{2016}\natexlab{}.
\newblock \showarticletitle{Fused-layer CNN accelerators}. In \bibinfo{booktitle}{\emph{2016 49th Annual IEEE/ACM International Symposium on Microarchitecture (MICRO)}}. \bibinfo{pages}{1--12}.
\newblock
\urldef\tempurl%
\url{https://doi.org/10.1109/MICRO.2016.7783725}
\showDOI{\tempurl}


\bibitem[Arora et~al\mbox{.}(2015)]%
        {static}
\bibfield{author}{\bibinfo{person}{Manish Arora}, \bibinfo{person}{Srilatha Manne}, \bibinfo{person}{Indrani Paul}, \bibinfo{person}{Nuwan Jayasena}, {and} \bibinfo{person}{Dean~M. Tullsen}.} \bibinfo{year}{2015}\natexlab{}.
\newblock \showarticletitle{Understanding idle behavior and power gating mechanisms in the context of modern benchmarks on CPU-GPU Integrated systems}. In \bibinfo{booktitle}{\emph{2015 IEEE 21st International Symposium on High Performance Computer Architecture (HPCA)}}. \bibinfo{pages}{366--377}.
\newblock
\urldef\tempurl%
\url{https://doi.org/10.1109/HPCA.2015.7056047}
\showDOI{\tempurl}


\bibitem[Barroso et~al\mbox{.}(2013)]%
        {41606}
\bibfield{author}{\bibinfo{person}{Luiz~André Barroso}, \bibinfo{person}{Jimmy Clidaras}, {and} \bibinfo{person}{Urs Hölzle}.} \bibinfo{year}{2013}\natexlab{}.
\newblock \bibinfo{booktitle}{\emph{The Datacenter as a Computer: An Introduction to the Design of Warehouse-Scale Machines, Second Edition}}.
\newblock
\urldef\tempurl%
\url{http://dx.doi.org/10.2200/S00516ED2V01Y201306CAC024}
\showURL{%
\tempurl}


\bibitem[Brown et~al\mbox{.}(2020)]%
        {brown2020languagemodelsfewshotlearners}
\bibfield{author}{\bibinfo{person}{Tom~B. Brown}, \bibinfo{person}{Benjamin Mann}, \bibinfo{person}{Nick Ryder}, \bibinfo{person}{Melanie Subbiah}, \bibinfo{person}{Jared Kaplan}, \bibinfo{person}{Prafulla Dhariwal}, \bibinfo{person}{Arvind Neelakantan}, \bibinfo{person}{Pranav Shyam}, \bibinfo{person}{Girish Sastry}, \bibinfo{person}{Amanda Askell}, \bibinfo{person}{Sandhini Agarwal}, \bibinfo{person}{Ariel Herbert-Voss}, \bibinfo{person}{Gretchen Krueger}, \bibinfo{person}{Tom Henighan}, \bibinfo{person}{Rewon Child}, \bibinfo{person}{Aditya Ramesh}, \bibinfo{person}{Daniel~M. Ziegler}, \bibinfo{person}{Jeffrey Wu}, \bibinfo{person}{Clemens Winter}, \bibinfo{person}{Christopher Hesse}, \bibinfo{person}{Mark Chen}, \bibinfo{person}{Eric Sigler}, \bibinfo{person}{Mateusz Litwin}, \bibinfo{person}{Scott Gray}, \bibinfo{person}{Benjamin Chess}, \bibinfo{person}{Jack Clark}, \bibinfo{person}{Christopher Berner}, \bibinfo{person}{Sam McCandlish}, \bibinfo{person}{Alec Radford}, \bibinfo{person}{Ilya Sutskever},
  {and} \bibinfo{person}{Dario Amodei}.} \bibinfo{year}{2020}\natexlab{}.
\newblock \bibinfo{title}{Language Models are Few-Shot Learners}.
\newblock
\newblock
\showeprint[arxiv]{2005.14165}~[cs.CL]
\urldef\tempurl%
\url{https://arxiv.org/abs/2005.14165}
\showURL{%
\tempurl}


\bibitem[Cai et~al\mbox{.}(2025)]%
        {soma}
\bibfield{author}{\bibinfo{person}{Jingwei Cai}, \bibinfo{person}{Xuan Wang}, \bibinfo{person}{Mingyu Gao}, \bibinfo{person}{Sen Peng}, \bibinfo{person}{Zijian Zhu}, \bibinfo{person}{Yuchen Wei}, \bibinfo{person}{Zuotong Wu}, {and} \bibinfo{person}{Kaisheng Ma}.} \bibinfo{year}{2025}\natexlab{}.
\newblock \showarticletitle{SoMA: Identifying, Exploring, and Understanding the DRAM Communication Scheduling Space for DNN Accelerators}. In \bibinfo{booktitle}{\emph{31st Symposium on High Performance Computer Architecture (HPCA)}} (Las Vegas, NV, USA). \bibinfo{pages}{533--548}.
\newblock


\bibitem[Cai et~al\mbox{.}(2023)]%
        {set}
\bibfield{author}{\bibinfo{person}{Jingwei Cai}, \bibinfo{person}{Yuchen Wei}, \bibinfo{person}{Zuotong Wu}, \bibinfo{person}{Sen Peng}, {and} \bibinfo{person}{Kaisheng Ma}.} \bibinfo{year}{2023}\natexlab{}.
\newblock \showarticletitle{Inter-layer Scheduling Space Definition and Exploration for Tiled Accelerators}. In \bibinfo{booktitle}{\emph{Proceedings of the 50th Annual International Symposium on Computer Architecture}} (Orlando, FL, USA) \emph{(\bibinfo{series}{ISCA '23})}. \bibinfo{publisher}{Association for Computing Machinery}, \bibinfo{address}{New York, NY, USA}, Article \bibinfo{articleno}{13}, \bibinfo{numpages}{17}~pages.
\newblock
\showISBNx{9798400700958}
\urldef\tempurl%
\url{https://doi.org/10.1145/3579371.3589048}
\showDOI{\tempurl}


\bibitem[Cai et~al\mbox{.}(2024)]%
        {gemini}
\bibfield{author}{\bibinfo{person}{Jingwei Cai}, \bibinfo{person}{Zuotong Wu}, \bibinfo{person}{Sen Peng}, \bibinfo{person}{Yuchen Wei}, \bibinfo{person}{Zhanhong Tan}, \bibinfo{person}{Guiming Shi}, \bibinfo{person}{Mingyu Gao}, {and} \bibinfo{person}{Kaisheng Ma}.} \bibinfo{year}{2024}\natexlab{}.
\newblock \showarticletitle{Gemini: Mapping and Architecture Co-exploration for Large-scale DNN Chiplet Accelerators}. In \bibinfo{booktitle}{\emph{30th Symposium on High Performance Computer Architecture (HPCA)}} (Edinburgh, Scotland). \bibinfo{pages}{156--171}.
\newblock


\bibitem[Chan(1996)]%
        {Convex2}
\bibfield{author}{\bibinfo{person}{Timothy~M Chan}.} \bibinfo{year}{1996}\natexlab{}.
\newblock \showarticletitle{Optimal output-sensitive convex hull algorithms in two and three dimensions}.
\newblock \bibinfo{journal}{\emph{Discrete \& computational geometry}} \bibinfo{volume}{16}, \bibinfo{number}{4} (\bibinfo{year}{1996}), \bibinfo{pages}{361--368}.
\newblock


\bibitem[Chen et~al\mbox{.}(2012)]%
        {Catci_3dd}
\bibfield{author}{\bibinfo{person}{Ke Chen}, \bibinfo{person}{Sheng Li}, \bibinfo{person}{Naveen Muralimanohar}, \bibinfo{person}{Jung~Ho Ahn}, \bibinfo{person}{Jay~B. Brockman}, {and} \bibinfo{person}{Norman~P. Jouppi}.} \bibinfo{year}{2012}\natexlab{}.
\newblock \showarticletitle{CACTI-3DD: Architecture-level modeling for 3D die-stacked DRAM main memory}. In \bibinfo{booktitle}{\emph{2012 Design, Automation \& Test in Europe Conference \& Exhibition (DATE)}}. \bibinfo{pages}{33--38}.
\newblock
\urldef\tempurl%
\url{https://doi.org/10.1109/DATE.2012.6176428}
\showDOI{\tempurl}


\bibitem[Chen et~al\mbox{.}(2019)]%
        {eyeriss_v2}
\bibfield{author}{\bibinfo{person}{Yu-Hsin Chen}, \bibinfo{person}{Tushar Krishna}, \bibinfo{person}{Joel Emer}, {and} \bibinfo{person}{Vivienne Sze}.} \bibinfo{year}{2019}\natexlab{}.
\newblock \showarticletitle{Eyeriss v2: A Flexible Accelerator for Emerging Deep Neural Networks on Mobile Devices}.
\newblock \bibinfo{journal}{\emph{IEEE Journal on Emerging and Selected Topics in Circuits and Systems (JETCAS)}} \bibinfo{volume}{9}, \bibinfo{number}{2} (\bibinfo{year}{2019}), \bibinfo{pages}{292--308}.
\newblock


\bibitem[Chen et~al\mbox{.}(2017)]%
        {eyeriss}
\bibfield{author}{\bibinfo{person}{Yu-Hsin Chen}, \bibinfo{person}{Tushar Krishna}, \bibinfo{person}{Joel~S. Emer}, {and} \bibinfo{person}{Vivienne Sze}.} \bibinfo{year}{2017}\natexlab{}.
\newblock \showarticletitle{Eyeriss: An Energy-Efficient Reconfigurable Accelerator for Deep Convolutional Neural Networks}.
\newblock \bibinfo{journal}{\emph{IEEE Journal of Solid-State Circuits}} \bibinfo{volume}{52}, \bibinfo{number}{1} (\bibinfo{year}{2017}), \bibinfo{pages}{127--138}.
\newblock
\urldef\tempurl%
\url{https://doi.org/10.1109/JSSC.2016.2616357}
\showDOI{\tempurl}


\bibitem[Das~Sharma(2022)]%
        {UCIe_}
\bibfield{author}{\bibinfo{person}{Debendra Das~Sharma}.} \bibinfo{year}{2022}\natexlab{}.
\newblock \bibinfo{title}{{Universal Chiplet Interconnect Express (UCIe): Building an Open Chiplet Ecosystem}}.
\newblock \bibinfo{howpublished}{UCIe Consortium White Paper}.
\newblock
\urldef\tempurl%
\url{https://www.uciexpress.org/_files/ugd/0c1418_c5970a68ab214ffc97fab16d11581449.pdf}
\showURL{%
\tempurl}


\bibitem[Das~Sharma et~al\mbox{.}(2022)]%
        {UCIe}
\bibfield{author}{\bibinfo{person}{Debendra Das~Sharma}, \bibinfo{person}{Gerald Pasdast}, \bibinfo{person}{Zhiguo Qian}, {and} \bibinfo{person}{Kemal Aygun}.} \bibinfo{year}{2022}\natexlab{}.
\newblock \showarticletitle{Universal Chiplet Interconnect Express (UCIe): An Open Industry Standard for Innovations With Chiplets at Package Level}.
\newblock \bibinfo{journal}{\emph{IEEE Transactions on Components, Packaging and Manufacturing Technology}} \bibinfo{volume}{12}, \bibinfo{number}{9} (\bibinfo{year}{2022}), \bibinfo{pages}{1423--1431}.
\newblock
\urldef\tempurl%
\url{https://doi.org/10.1109/TCPMT.2022.3207195}
\showDOI{\tempurl}


\bibitem[Dave et~al\mbox{.}(2024)]%
        {explainable_dse}
\bibfield{author}{\bibinfo{person}{Shail Dave}, \bibinfo{person}{Tony Nowatzki}, {and} \bibinfo{person}{Aviral Shrivastava}.} \bibinfo{year}{2024}\natexlab{}.
\newblock \showarticletitle{Explainable-DSE: An Agile and Explainable Exploration of Efficient HW/SW Codesigns of Deep Learning Accelerators Using Bottleneck Analysis}. In \bibinfo{booktitle}{\emph{Proceedings of the 28th ACM International Conference on Architectural Support for Programming Languages and Operating Systems, Volume 4}} (Vancouver, BC, Canada) \emph{(\bibinfo{series}{ASPLOS '23})}. \bibinfo{publisher}{Association for Computing Machinery}, \bibinfo{address}{New York, NY, USA}, \bibinfo{pages}{87–107}.
\newblock
\showISBNx{9798400703942}
\urldef\tempurl%
\url{https://doi.org/10.1145/3623278.3624772}
\showDOI{\tempurl}


\bibitem[Ding et~al\mbox{.}(2022)]%
        {Scaling_Up_Your_Kernels_to_31x31}
\bibfield{author}{\bibinfo{person}{Xiaohan Ding}, \bibinfo{person}{Xiangyu Zhang}, \bibinfo{person}{Yizhuang Zhou}, \bibinfo{person}{Jungong Han}, \bibinfo{person}{Guiguang Ding}, {and} \bibinfo{person}{Jian Sun}.} \bibinfo{year}{2022}\natexlab{}.
\newblock \bibinfo{title}{Scaling Up Your Kernels to 31x31: Revisiting Large Kernel Design in CNNs}.
\newblock
\newblock
\showeprint[arxiv]{2203.06717}~[cs.CV]
\urldef\tempurl%
\url{https://arxiv.org/abs/2203.06717}
\showURL{%
\tempurl}


\bibitem[Dosovitskiy et~al\mbox{.}(2021)]%
        {vit}
\bibfield{author}{\bibinfo{person}{Alexey Dosovitskiy}, \bibinfo{person}{Lucas Beyer}, \bibinfo{person}{Alexander Kolesnikov}, \bibinfo{person}{Dirk Weissenborn}, \bibinfo{person}{Xiaohua Zhai}, \bibinfo{person}{Thomas Unterthiner}, \bibinfo{person}{Mostafa Dehghani}, \bibinfo{person}{Matthias Minderer}, \bibinfo{person}{Georg Heigold}, \bibinfo{person}{Sylvain Gelly}, \bibinfo{person}{Jakob Uszkoreit}, {and} \bibinfo{person}{Neil Houlsby}.} \bibinfo{year}{2021}\natexlab{}.
\newblock \bibinfo{title}{An Image is Worth 16x16 Words: Transformers for Image Recognition at Scale}.
\newblock
\newblock
\showeprint[arxiv]{2010.11929}~[cs.CV]
\urldef\tempurl%
\url{https://arxiv.org/abs/2010.11929}
\showURL{%
\tempurl}


\bibitem[Du et~al\mbox{.}(2015)]%
        {shidiannao}
\bibfield{author}{\bibinfo{person}{Zidong Du}, \bibinfo{person}{Robert Fasthuber}, \bibinfo{person}{Tianshi Chen}, \bibinfo{person}{Paolo Ienne}, \bibinfo{person}{Ling Li}, \bibinfo{person}{Tao Luo}, \bibinfo{person}{Xiaobing Feng}, \bibinfo{person}{Yunji Chen}, {and} \bibinfo{person}{Olivier Temam}.} \bibinfo{year}{2015}\natexlab{}.
\newblock \showarticletitle{ShiDianNao: Shifting vision processing closer to the sensor}. In \bibinfo{booktitle}{\emph{2015 ACM/IEEE 42nd Annual International Symposium on Computer Architecture (ISCA)}}. \bibinfo{pages}{92--104}.
\newblock
\urldef\tempurl%
\url{https://doi.org/10.1145/2749469.2750389}
\showDOI{\tempurl}


\bibitem[Feng and Ma(2022)]%
        {Chiplet_actuary}
\bibfield{author}{\bibinfo{person}{Yinxiao Feng} {and} \bibinfo{person}{Kaisheng Ma}.} \bibinfo{year}{2022}\natexlab{}.
\newblock \showarticletitle{Chiplet actuary: a quantitative cost model and multi-chiplet architecture exploration}. In \bibinfo{booktitle}{\emph{Proceedings of the 59th ACM/IEEE Design Automation Conference}} (San Francisco, California) \emph{(\bibinfo{series}{DAC '22})}. \bibinfo{publisher}{Association for Computing Machinery}, \bibinfo{address}{New York, NY, USA}, \bibinfo{pages}{121–126}.
\newblock
\showISBNx{9781450391429}
\urldef\tempurl%
\url{https://doi.org/10.1145/3489517.3530428}
\showDOI{\tempurl}


\bibitem[Flautner et~al\mbox{.}(2002)]%
        {drowsy}
\bibfield{author}{\bibinfo{person}{Kriszti\'{a}n Flautner}, \bibinfo{person}{Nam~Sung Kim}, \bibinfo{person}{Steve Martin}, \bibinfo{person}{David Blaauw}, {and} \bibinfo{person}{Trevor Mudge}.} \bibinfo{year}{2002}\natexlab{}.
\newblock \showarticletitle{Drowsy caches: simple techniques for reducing leakage power}. In \bibinfo{booktitle}{\emph{Proceedings of the 29th Annual International Symposium on Computer Architecture}} (Anchorage, Alaska) \emph{(\bibinfo{series}{ISCA '02})}. \bibinfo{publisher}{IEEE Computer Society}, \bibinfo{address}{USA}, \bibinfo{pages}{148–157}.
\newblock
\showISBNx{076951605X}


\bibitem[Genc et~al\mbox{.}(2024)]%
        {stellar}
\bibfield{author}{\bibinfo{person}{Hasan~Nazim Genc}, \bibinfo{person}{Hansung Kim}, \bibinfo{person}{Prashanth Ganesh}, {and} \bibinfo{person}{Yakun~Sophia Shao}.} \bibinfo{year}{2024}\natexlab{}.
\newblock \showarticletitle{Stellar: An Automated Design Framework for Dense and Sparse Spatial Accelerators}. In \bibinfo{booktitle}{\emph{2024 57th IEEE/ACM International Symposium on Microarchitecture (MICRO)}}. \bibinfo{pages}{409--422}.
\newblock
\urldef\tempurl%
\url{https://doi.org/10.1109/MICRO61859.2024.00038}
\showDOI{\tempurl}


\bibitem[Glint et~al\mbox{.}(2024)]%
        {deepfrack}
\bibfield{author}{\bibinfo{person}{Tom Glint}, \bibinfo{person}{Mithil Pechimuthu}, {and} \bibinfo{person}{Joycee Mekie}.} \bibinfo{year}{2024}\natexlab{}.
\newblock \showarticletitle{DeepFrack: A Comprehensive Framework for Layer Fusion, Face Tiling, and Efficient Mapping in DNN Hardware Accelerators}. In \bibinfo{booktitle}{\emph{2024 Design, Automation \& Test in Europe Conference \& Exhibition (DATE)}}. \bibinfo{pages}{1--6}.
\newblock
\urldef\tempurl%
\url{https://doi.org/10.23919/DATE58400.2024.10546624}
\showDOI{\tempurl}


\bibitem[Goodfellow et~al\mbox{.}(2020)]%
        {gan}
\bibfield{author}{\bibinfo{person}{Ian Goodfellow}, \bibinfo{person}{Jean Pouget-Abadie}, \bibinfo{person}{Mehdi Mirza}, \bibinfo{person}{Bing Xu}, \bibinfo{person}{David Warde-Farley}, \bibinfo{person}{Sherjil Ozair}, \bibinfo{person}{Aaron Courville}, {and} \bibinfo{person}{Yoshua Bengio}.} \bibinfo{year}{2020}\natexlab{}.
\newblock \showarticletitle{Generative adversarial networks}.
\newblock \bibinfo{journal}{\emph{Commun. ACM}} \bibinfo{volume}{63}, \bibinfo{number}{11} (\bibinfo{year}{2020}), \bibinfo{pages}{139--144}.
\newblock


\bibitem[Graening et~al\mbox{.}(2023)]%
        {how_small_is_too_small}
\bibfield{author}{\bibinfo{person}{Alexander Graening}, \bibinfo{person}{Saptadeep Pal}, {and} \bibinfo{person}{Puneet Gupta}.} \bibinfo{year}{2023}\natexlab{}.
\newblock \showarticletitle{Chiplets: How Small is too Small?}. In \bibinfo{booktitle}{\emph{2023 60th ACM/IEEE Design Automation Conference (DAC)}}. \bibinfo{pages}{1--6}.
\newblock
\urldef\tempurl%
\url{https://doi.org/10.1109/DAC56929.2023.10247947}
\showDOI{\tempurl}


\bibitem[Graening et~al\mbox{.}(2025)]%
        {catch}
\bibfield{author}{\bibinfo{person}{Alexander Graening}, \bibinfo{person}{Jonti Talukdar}, \bibinfo{person}{Saptadeep Pal}, \bibinfo{person}{Krishnendu Chakrabarty}, {and} \bibinfo{person}{Puneet Gupta}.} \bibinfo{year}{2025}\natexlab{}.
\newblock \showarticletitle{CATCH: a Cost Analysis Tool for Co-optimization of Chiplet-based Heterogeneous Systems}.
\newblock \bibinfo{journal}{\emph{arXiv preprint arXiv:2503.15753}} (\bibinfo{year}{2025}).
\newblock


\bibitem[Hao et~al\mbox{.}(2023)]%
        {Monad}
\bibfield{author}{\bibinfo{person}{Xiaochen Hao}, \bibinfo{person}{Zijian Ding}, \bibinfo{person}{Jieming Yin}, \bibinfo{person}{Yuan Wang}, {and} \bibinfo{person}{Yun Liang}.} \bibinfo{year}{2023}\natexlab{}.
\newblock \showarticletitle{Monad: Towards Cost-Effective Specialization for Chiplet-Based Spatial Accelerators}. In \bibinfo{booktitle}{\emph{2023 IEEE/ACM International Conference on Computer Aided Design (ICCAD)}}. \bibinfo{pages}{1--9}.
\newblock
\urldef\tempurl%
\url{https://doi.org/10.1109/ICCAD57390.2023.10323880}
\showDOI{\tempurl}


\bibitem[Hong et~al\mbox{.}(2023)]%
        {dosa}
\bibfield{author}{\bibinfo{person}{Charles Hong}, \bibinfo{person}{Qijing Huang}, \bibinfo{person}{Grace Dinh}, \bibinfo{person}{Mahesh Subedar}, {and} \bibinfo{person}{Yakun~Sophia Shao}.} \bibinfo{year}{2023}\natexlab{}.
\newblock \showarticletitle{DOSA: Differentiable Model-Based One-Loop Search for DNN Accelerators}. In \bibinfo{booktitle}{\emph{2023 56th IEEE/ACM International Symposium on Microarchitecture (MICRO)}}. \bibinfo{pages}{209--224}.
\newblock


\bibitem[Huang et~al\mbox{.}(2021)]%
        {cosa}
\bibfield{author}{\bibinfo{person}{Qijing Huang}, \bibinfo{person}{Minwoo Kang}, \bibinfo{person}{Grace Dinh}, \bibinfo{person}{Thomas Norell}, \bibinfo{person}{Aravind Kalaiah}, \bibinfo{person}{James Demmel}, \bibinfo{person}{John Wawrzynek}, {and} \bibinfo{person}{Yakun~Sophia Shao}.} \bibinfo{year}{2021}\natexlab{}.
\newblock \showarticletitle{CoSA: Scheduling by Constrained Optimization for Spatial Accelerators}. In \bibinfo{booktitle}{\emph{2021 ACM/IEEE 48th Annual International Symposium on Computer Architecture (ISCA)}}. \bibinfo{pages}{554--566}.
\newblock
\urldef\tempurl%
\url{https://doi.org/10.1109/ISCA52012.2021.00050}
\showDOI{\tempurl}


\bibitem[Huang et~al\mbox{.}(2024)]%
        {mind_the_gap}
\bibfield{author}{\bibinfo{person}{Qijing Huang}, \bibinfo{person}{Po-An Tsai}, \bibinfo{person}{Joel~S. Emer}, {and} \bibinfo{person}{Angshuman Parashar}.} \bibinfo{year}{2024}\natexlab{}.
\newblock \showarticletitle{Mind the Gap: Attainable Data Movement and Operational Intensity Bounds for Tensor Algorithms}. In \bibinfo{booktitle}{\emph{2024 ACM/IEEE 51st Annual International Symposium on Computer Architecture (ISCA)}}. \bibinfo{pages}{150--166}.
\newblock
\urldef\tempurl%
\url{https://doi.org/10.1109/ISCA59077.2024.00021}
\showDOI{\tempurl}


\bibitem[{JEDEC Solid State Technology Association}(2022)]%
        {jedec_hbm3_2022}
\bibfield{author}{\bibinfo{person}{{JEDEC Solid State Technology Association}}.} \bibinfo{year}{2022}\natexlab{}.
\newblock \bibinfo{title}{{JEDEC Publishes HBM3 Update to High Bandwidth Memory (HBM) Standard}}.
\newblock \bibinfo{howpublished}{\url{https://www.jedec.org/news/pressreleases/jedec-publishes-hbm3-update-high-bandwidth-memory-hbm-standard}}.
\newblock
\newblock
\shownote{Press Release}.


\bibitem[Jouppi et~al\mbox{.}(2012)]%
        {Catci_io}
\bibfield{author}{\bibinfo{person}{Norman~P. Jouppi}, \bibinfo{person}{Andrew~B. Kahng}, \bibinfo{person}{Naveen Muralimanohar}, {and} \bibinfo{person}{Vaishnav Srinivas}.} \bibinfo{year}{2012}\natexlab{}.
\newblock \showarticletitle{CACTI-IO: CACTI with off-chip power-area-timing models}. In \bibinfo{booktitle}{\emph{2012 IEEE/ACM International Conference on Computer-Aided Design (ICCAD)}}. \bibinfo{pages}{294--301}.
\newblock


\bibitem[Kao et~al\mbox{.}(2023)]%
        {flat}
\bibfield{author}{\bibinfo{person}{Sheng-Chun Kao}, \bibinfo{person}{Suvinay Subramanian}, \bibinfo{person}{Gaurav Agrawal}, \bibinfo{person}{Amir Yazdanbakhsh}, {and} \bibinfo{person}{Tushar Krishna}.} \bibinfo{year}{2023}\natexlab{}.
\newblock \showarticletitle{FLAT: An Optimized Dataflow for Mitigating Attention Bottlenecks}. In \bibinfo{booktitle}{\emph{Proceedings of the 28th ACM International Conference on Architectural Support for Programming Languages and Operating Systems, Volume 2}} (Vancouver, BC, Canada) \emph{(\bibinfo{series}{ASPLOS 2023})}. \bibinfo{publisher}{Association for Computing Machinery}, \bibinfo{address}{New York, NY, USA}, \bibinfo{pages}{295–310}.
\newblock
\showISBNx{9781450399166}
\urldef\tempurl%
\url{https://doi.org/10.1145/3575693.3575747}
\showDOI{\tempurl}


\bibitem[Ko et~al\mbox{.}(2024)]%
        {dfmodel}
\bibfield{author}{\bibinfo{person}{Sho Ko}, \bibinfo{person}{Nathan Zhang}, \bibinfo{person}{Olivia Hsu}, \bibinfo{person}{Ardavan Pedram}, {and} \bibinfo{person}{Kunle Olukotun}.} \bibinfo{year}{2024}\natexlab{}.
\newblock \bibinfo{title}{DFModel: Design Space Optimization of Large-Scale Systems Exploiting Dataflow Mappings}.
\newblock
\newblock
\showeprint[arxiv]{2412.16432}~[cs.AR]
\urldef\tempurl%
\url{https://arxiv.org/abs/2412.16432}
\showURL{%
\tempurl}


\bibitem[Krishnan et~al\mbox{.}(2022)]%
        {BigLittle}
\bibfield{author}{\bibinfo{person}{Gokul Krishnan}, \bibinfo{person}{A.~Alper Goksoy}, \bibinfo{person}{Sumit~K. Mandal}, \bibinfo{person}{Zhenyu Wang}, \bibinfo{person}{Chaitali Chakrabarti}, \bibinfo{person}{Jae-sun Seo}, \bibinfo{person}{Umit~Y. Ogras}, {and} \bibinfo{person}{Yu Cao}.} \bibinfo{year}{2022}\natexlab{}.
\newblock \showarticletitle{{Big-Little Chiplets for In-Memory Acceleration of DNNs: A Scalable Heterogeneous Architecture}}. In \bibinfo{booktitle}{\emph{Proc. 41st IEEE/ACM International Conference on Computer-Aided Design (ICCAD)}}.
\newblock
\urldef\tempurl%
\url{https://doi.org/10.1145/3508352.3549447}
\showDOI{\tempurl}


\bibitem[Krizhevsky et~al\mbox{.}(2012)]%
        {alexnet}
\bibfield{author}{\bibinfo{person}{Alex Krizhevsky}, \bibinfo{person}{Ilya Sutskever}, {and} \bibinfo{person}{Geoffrey~E Hinton}.} \bibinfo{year}{2012}\natexlab{}.
\newblock \showarticletitle{Imagenet classification with deep convolutional neural networks}.
\newblock \bibinfo{journal}{\emph{Advances in neural information processing systems}}  \bibinfo{volume}{25} (\bibinfo{year}{2012}).
\newblock


\bibitem[Kwon et~al\mbox{.}(2020)]%
        {maestro}
\bibfield{author}{\bibinfo{person}{Hyoukjun Kwon}, \bibinfo{person}{Prasanth Chatarasi}, \bibinfo{person}{Vivek Sarkar}, \bibinfo{person}{Tushar Krishna}, \bibinfo{person}{Michael Pellauer}, {and} \bibinfo{person}{Angshuman Parashar}.} \bibinfo{year}{2020}\natexlab{}.
\newblock \showarticletitle{MAESTRO: A Data-Centric Approach to Understand Reuse, Performance, and Hardware Cost of DNN Mappings}.
\newblock \bibinfo{journal}{\emph{IEEE Micro}} \bibinfo{volume}{40}, \bibinfo{number}{3} (\bibinfo{year}{2020}), \bibinfo{pages}{20--29}.
\newblock
\urldef\tempurl%
\url{https://doi.org/10.1109/MM.2020.2985963}
\showDOI{\tempurl}


\bibitem[Kwon et~al\mbox{.}(2021)]%
        {heterdata}
\bibfield{author}{\bibinfo{person}{Hyoukjun Kwon}, \bibinfo{person}{Liangzhen Lai}, \bibinfo{person}{Michael Pellauer}, \bibinfo{person}{Tushar Krishna}, \bibinfo{person}{Yu-Hsin Chen}, {and} \bibinfo{person}{Vikas Chandra}.} \bibinfo{year}{2021}\natexlab{}.
\newblock \showarticletitle{Heterogeneous dataflow accelerators for multi-DNN workloads}. In \bibinfo{booktitle}{\emph{2021 IEEE International Symposium on High-Performance Computer Architecture (HPCA)}}. IEEE, \bibinfo{pages}{71--83}.
\newblock


\bibitem[Leviathan et~al\mbox{.}(2023)]%
        {sd}
\bibfield{author}{\bibinfo{person}{Yaniv Leviathan}, \bibinfo{person}{Matan Kalman}, {and} \bibinfo{person}{Yossi Matias}.} \bibinfo{year}{2023}\natexlab{}.
\newblock \bibinfo{title}{Fast Inference from Transformers via Speculative Decoding}.
\newblock
\newblock
\showeprint[arxiv]{2211.17192}~[cs.LG]
\urldef\tempurl%
\url{https://arxiv.org/abs/2211.17192}
\showURL{%
\tempurl}


\bibitem[Li and Wentzlaff(2024)]%
        {lucie}
\bibfield{author}{\bibinfo{person}{Zixi Li} {and} \bibinfo{person}{David Wentzlaff}.} \bibinfo{year}{2024}\natexlab{}.
\newblock \showarticletitle{LUCIE: A Universal Chiplet-Interposer Design Framework for Plug-and-Play Integration}. In \bibinfo{booktitle}{\emph{2024 57th IEEE/ACM International Symposium on Microarchitecture (MICRO)}}. \bibinfo{pages}{423--436}.
\newblock
\urldef\tempurl%
\url{https://doi.org/10.1109/MICRO61859.2024.00039}
\showDOI{\tempurl}


\bibitem[Lin et~al\mbox{.}(2018)]%
        {Constraints}
\bibfield{author}{\bibinfo{person}{Shih-Chieh Lin}, \bibinfo{person}{Yunqi Zhang}, \bibinfo{person}{Chang-Hong Hsu}, \bibinfo{person}{Matt Skach}, \bibinfo{person}{Md~E Haque}, \bibinfo{person}{Lingjia Tang}, {and} \bibinfo{person}{Jason Mars}.} \bibinfo{year}{2018}\natexlab{}.
\newblock \showarticletitle{The architectural implications of autonomous driving: Constraints and acceleration}. In \bibinfo{booktitle}{\emph{Proceedings of the twenty-third international conference on architectural support for programming languages and operating systems}}. \bibinfo{pages}{751--766}.
\newblock


\bibitem[Mei et~al\mbox{.}(2023)]%
        {defines}
\bibfield{author}{\bibinfo{person}{Linyan Mei}, \bibinfo{person}{Koen Goetschalckx}, \bibinfo{person}{Arne Symons}, {and} \bibinfo{person}{Marian Verhelst}.} \bibinfo{year}{2023}\natexlab{}.
\newblock \showarticletitle{DeFiNES: Enabling Fast Exploration of the Depth-first Scheduling Space for DNN Accelerators through Analytical Modeling}. In \bibinfo{booktitle}{\emph{2023 IEEE International Symposium on High-Performance Computer Architecture (HPCA)}}. \bibinfo{pages}{570--583}.
\newblock
\urldef\tempurl%
\url{https://doi.org/10.1109/HPCA56546.2023.10071098}
\showDOI{\tempurl}


\bibitem[Mittal and Vetter(2014)]%
        {gpuenergy}
\bibfield{author}{\bibinfo{person}{Sparsh Mittal} {and} \bibinfo{person}{Jeffrey~S. Vetter}.} \bibinfo{year}{2014}\natexlab{}.
\newblock \showarticletitle{A Survey of Methods for Analyzing and Improving GPU Energy Efficiency}.
\newblock \bibinfo{journal}{\emph{ACM Comput. Surv.}} \bibinfo{volume}{47}, \bibinfo{number}{2}, Article \bibinfo{articleno}{19} (\bibinfo{date}{Aug.} \bibinfo{year}{2014}), \bibinfo{numpages}{23}~pages.
\newblock
\showISSN{0360-0300}
\urldef\tempurl%
\url{https://doi.org/10.1145/2636342}
\showDOI{\tempurl}


\bibitem[Naffziger et~al\mbox{.}(2021)]%
        {pioneering_amd}
\bibfield{author}{\bibinfo{person}{Samuel Naffziger}, \bibinfo{person}{Noah Beck}, \bibinfo{person}{Thomas Burd}, \bibinfo{person}{Kevin Lepak}, \bibinfo{person}{Gabriel~H. Loh}, \bibinfo{person}{Mahesh Subramony}, {and} \bibinfo{person}{Sean White}.} \bibinfo{year}{2021}\natexlab{}.
\newblock \showarticletitle{Pioneering Chiplet Technology and Design for the AMD EPYC™ and Ryzen™ Processor Families : Industrial Product}. In \bibinfo{booktitle}{\emph{2021 ACM/IEEE 48th Annual International Symposium on Computer Architecture (ISCA)}}. \bibinfo{pages}{57--70}.
\newblock
\urldef\tempurl%
\url{https://doi.org/10.1109/ISCA52012.2021.00014}
\showDOI{\tempurl}


\bibitem[Nayak et~al\mbox{.}(2024)]%
        {fusemax}
\bibfield{author}{\bibinfo{person}{Nandeeka Nayak}, \bibinfo{person}{Xinrui Wu}, \bibinfo{person}{Toluwanimi~O. Odemuyiwa}, \bibinfo{person}{Michael Pellauer}, \bibinfo{person}{Joel~S. Emer}, {and} \bibinfo{person}{Christopher~W. Fletcher}.} \bibinfo{year}{2024}\natexlab{}.
\newblock \showarticletitle{FuseMax: Leveraging Extended Einsums to Optimize Attention Accelerator Design}. In \bibinfo{booktitle}{\emph{2024 57th IEEE/ACM International Symposium on Microarchitecture (MICRO)}}. \bibinfo{pages}{1458--1473}.
\newblock
\urldef\tempurl%
\url{https://doi.org/10.1109/MICRO61859.2024.00107}
\showDOI{\tempurl}


\bibitem[Odema et~al\mbox{.}(2024)]%
        {scar}
\bibfield{author}{\bibinfo{person}{Mohanad Odema}, \bibinfo{person}{Luke Chen}, \bibinfo{person}{Hyoukjun Kwon}, {and} \bibinfo{person}{Mohammad~Abdullah Al~Faruque}.} \bibinfo{year}{2024}\natexlab{}.
\newblock \showarticletitle{SCAR: Scheduling Multi-Model AI Workloads on Heterogeneous Multi-Chiplet Module Accelerators}. In \bibinfo{booktitle}{\emph{2024 57th IEEE/ACM International Symposium on Microarchitecture (MICRO)}}. \bibinfo{pages}{565--579}.
\newblock
\urldef\tempurl%
\url{https://doi.org/10.1109/MICRO61859.2024.00049}
\showDOI{\tempurl}


\bibitem[Parashar et~al\mbox{.}(2019)]%
        {timeloop}
\bibfield{author}{\bibinfo{person}{Angshuman Parashar}, \bibinfo{person}{Priyanka Raina}, \bibinfo{person}{Yakun~Sophia Shao}, \bibinfo{person}{Yu-Hsin Chen}, \bibinfo{person}{Victor~A Ying}, \bibinfo{person}{Anurag Mukkara}, \bibinfo{person}{Rangharajan Venkatesan}, \bibinfo{person}{Brucek Khailany}, \bibinfo{person}{Stephen~W Keckler}, {and} \bibinfo{person}{Joel Emer}.} \bibinfo{year}{2019}\natexlab{}.
\newblock \showarticletitle{Timeloop: A systematic approach to dnn accelerator evaluation}. In \bibinfo{booktitle}{\emph{2019 IEEE international symposium on performance analysis of systems and software (ISPASS)}}. \bibinfo{pages}{304--315}.
\newblock


\bibitem[Patel et~al\mbox{.}(2024)]%
        {splitwiseefficientgenerativellm}
\bibfield{author}{\bibinfo{person}{Pratyush Patel}, \bibinfo{person}{Esha Choukse}, \bibinfo{person}{Chaojie Zhang}, \bibinfo{person}{Aashaka Shah}, \bibinfo{person}{Íñigo Goiri}, \bibinfo{person}{Saeed Maleki}, {and} \bibinfo{person}{Ricardo Bianchini}.} \bibinfo{year}{2024}\natexlab{}.
\newblock \bibinfo{title}{Splitwise: Efficient generative LLM inference using phase splitting}.
\newblock
\newblock
\showeprint[arxiv]{2311.18677}~[cs.AR]
\urldef\tempurl%
\url{https://arxiv.org/abs/2311.18677}
\showURL{%
\tempurl}


\bibitem[Redmon et~al\mbox{.}(2016)]%
        {yolo}
\bibfield{author}{\bibinfo{person}{Joseph Redmon}, \bibinfo{person}{Santosh Divvala}, \bibinfo{person}{Ross Girshick}, {and} \bibinfo{person}{Ali Farhadi}.} \bibinfo{year}{2016}\natexlab{}.
\newblock \showarticletitle{You only look once: Unified, real-time object detection}. In \bibinfo{booktitle}{\emph{Proceedings of the IEEE conference on computer vision and pattern recognition}}. \bibinfo{pages}{779--788}.
\newblock


\bibitem[Rombach et~al\mbox{.}(2022)]%
        {stable_diffusion}
\bibfield{author}{\bibinfo{person}{Robin Rombach}, \bibinfo{person}{Andreas Blattmann}, \bibinfo{person}{Dominik Lorenz}, \bibinfo{person}{Patrick Esser}, {and} \bibinfo{person}{Björn Ommer}.} \bibinfo{year}{2022}\natexlab{}.
\newblock \bibinfo{title}{High-Resolution Image Synthesis with Latent Diffusion Models}.
\newblock
\newblock
\showeprint[arxiv]{2112.10752}~[cs.CV]
\urldef\tempurl%
\url{https://arxiv.org/abs/2112.10752}
\showURL{%
\tempurl}


\bibitem[{Samsung Semiconductor}(2022)]%
        {samsung_k4z80325bc_datasheet}
\bibfield{author}{\bibinfo{person}{{Samsung Semiconductor}}.} \bibinfo{year}{2022}\natexlab{}.
\newblock \bibinfo{booktitle}{\emph{{K4Z80325BC-HC14 8Gb GDDR6 SDRAM Datasheet}}}.
\newblock \bibinfo{type}{Datasheet}. \bibinfo{institution}{Samsung Semiconductor}.
\newblock
\urldef\tempurl%
\url{https://datasheet.lcsc.com/lcsc/2204251615_Samsung-K4Z80325BC-HC14_C2920181.pdf}
\showURL{%
\tempurl}
\newblock
\shownote{8Gb GDDR6 256Mx32 Memory IC, Part Number: K4Z80325BC-HC14}.


\bibitem[Shao et~al\mbox{.}(2019)]%
        {simba}
\bibfield{author}{\bibinfo{person}{Yakun~Sophia Shao}, \bibinfo{person}{Jason Clemons}, \bibinfo{person}{Rangharajan Venkatesan}, \bibinfo{person}{Brian Zimmer}, \bibinfo{person}{Matthew Fojtik}, \bibinfo{person}{Nan Jiang}, \bibinfo{person}{Ben Keller}, \bibinfo{person}{Alicia Klinefelter}, \bibinfo{person}{Nathaniel Pinckney}, \bibinfo{person}{Priyanka Raina}, \bibinfo{person}{Stephen~G. Tell}, \bibinfo{person}{Yanqing Zhang}, \bibinfo{person}{William~J. Dally}, \bibinfo{person}{Joel Emer}, \bibinfo{person}{C.~Thomas Gray}, \bibinfo{person}{Brucek Khailany}, {and} \bibinfo{person}{Stephen~W. Keckler}.} \bibinfo{year}{2019}\natexlab{}.
\newblock \showarticletitle{Simba: Scaling Deep-Learning Inference with Multi-Chip-Module-Based Architecture}. In \bibinfo{booktitle}{\emph{Proceedings of the 52nd Annual IEEE/ACM International Symposium on Microarchitecture}} (Columbus, OH, USA) \emph{(\bibinfo{series}{MICRO '52})}. \bibinfo{publisher}{Association for Computing Machinery}, \bibinfo{address}{New York, NY, USA}, \bibinfo{pages}{14–27}.
\newblock
\showISBNx{9781450369381}
\urldef\tempurl%
\url{https://doi.org/10.1145/3352460.3358302}
\showDOI{\tempurl}


\bibitem[Sheng et~al\mbox{.}(2023)]%
        {sheng2023flexgenhighthroughputgenerativeinference}
\bibfield{author}{\bibinfo{person}{Ying Sheng}, \bibinfo{person}{Lianmin Zheng}, \bibinfo{person}{Binhang Yuan}, \bibinfo{person}{Zhuohan Li}, \bibinfo{person}{Max Ryabinin}, \bibinfo{person}{Daniel~Y. Fu}, \bibinfo{person}{Zhiqiang Xie}, \bibinfo{person}{Beidi Chen}, \bibinfo{person}{Clark Barrett}, \bibinfo{person}{Joseph~E. Gonzalez}, \bibinfo{person}{Percy Liang}, \bibinfo{person}{Christopher Ré}, \bibinfo{person}{Ion Stoica}, {and} \bibinfo{person}{Ce Zhang}.} \bibinfo{year}{2023}\natexlab{}.
\newblock \bibinfo{title}{FlexGen: High-Throughput Generative Inference of Large Language Models with a Single GPU}.
\newblock
\newblock
\showeprint[arxiv]{2303.06865}~[cs.LG]
\urldef\tempurl%
\url{https://arxiv.org/abs/2303.06865}
\showURL{%
\tempurl}


\bibitem[Silvano et~al\mbox{.}(2025)]%
        {silvano2025surveydeeplearninghardware}
\bibfield{author}{\bibinfo{person}{Cristina Silvano}, \bibinfo{person}{Daniele Ielmini}, \bibinfo{person}{Fabrizio Ferrandi}, \bibinfo{person}{Leandro Fiorin}, \bibinfo{person}{Serena Curzel}, \bibinfo{person}{Luca Benini}, \bibinfo{person}{Francesco Conti}, \bibinfo{person}{Angelo Garofalo}, \bibinfo{person}{Cristian Zambelli}, \bibinfo{person}{Enrico Calore}, \bibinfo{person}{Sebastiano~Fabio Schifano}, \bibinfo{person}{Maurizio Palesi}, \bibinfo{person}{Giuseppe Ascia}, \bibinfo{person}{Davide Patti}, \bibinfo{person}{Nicola Petra}, \bibinfo{person}{Davide~De Caro}, \bibinfo{person}{Luciano Lavagno}, \bibinfo{person}{Teodoro Urso}, \bibinfo{person}{Valeria Cardellini}, \bibinfo{person}{Gian~Carlo Cardarilli}, \bibinfo{person}{Robert Birke}, {and} \bibinfo{person}{Stefania Perri}.} \bibinfo{year}{2025}\natexlab{}.
\newblock \bibinfo{title}{A Survey on Deep Learning Hardware Accelerators for Heterogeneous HPC Platforms}.
\newblock
\newblock
\showeprint[arxiv]{2306.15552}~[cs.AR]
\urldef\tempurl%
\url{https://arxiv.org/abs/2306.15552}
\showURL{%
\tempurl}


\bibitem[Song et~al\mbox{.}(2019)]%
        {HyPar}
\bibfield{author}{\bibinfo{person}{Linghao Song}, \bibinfo{person}{Jiachen Mao}, \bibinfo{person}{Youwei Zhuo}, \bibinfo{person}{Xuehai Qian}, \bibinfo{person}{Hai Li}, {and} \bibinfo{person}{Yiran Chen}.} \bibinfo{year}{2019}\natexlab{}.
\newblock \showarticletitle{HyPar: Towards Hybrid Parallelism for Deep Learning Accelerator Array}. In \bibinfo{booktitle}{\emph{2019 IEEE International Symposium on High Performance Computer Architecture (HPCA)}}. \bibinfo{pages}{56--68}.
\newblock
\urldef\tempurl%
\url{https://doi.org/10.1109/HPCA.2019.00027}
\showDOI{\tempurl}


\bibitem[Tan et~al\mbox{.}(2020)]%
        {efficientdet}
\bibfield{author}{\bibinfo{person}{Mingxing Tan}, \bibinfo{person}{Ruoming Pang}, {and} \bibinfo{person}{Quoc~V Le}.} \bibinfo{year}{2020}\natexlab{}.
\newblock \showarticletitle{Efficientdet: Scalable and efficient object detection}. In \bibinfo{booktitle}{\emph{Proceedings of the IEEE/CVF conference on computer vision and pattern recognition}}. \bibinfo{pages}{10781--10790}.
\newblock


\bibitem[Tan et~al\mbox{.}(2024)]%
        {cocco}
\bibfield{author}{\bibinfo{person}{Zhanhong Tan}, \bibinfo{person}{Zijian Zhu}, {and} \bibinfo{person}{Kaisheng Ma}.} \bibinfo{year}{2024}\natexlab{}.
\newblock \showarticletitle{Cocco: Hardware-Mapping Co-Exploration towards Memory Capacity-Communication Optimization}. In \bibinfo{booktitle}{\emph{Proceedings of the 29th ACM International Conference on Architectural Support for Programming Languages and Operating Systems, Volume 1}} (La Jolla, CA, USA) \emph{(\bibinfo{series}{ASPLOS '24})}. \bibinfo{publisher}{Association for Computing Machinery}, \bibinfo{address}{New York, NY, USA}, \bibinfo{pages}{69–84}.
\newblock
\showISBNx{9798400703720}
\urldef\tempurl%
\url{https://doi.org/10.1145/3617232.3624865}
\showDOI{\tempurl}


\bibitem[Vaswani et~al\mbox{.}(2023)]%
        {vaswani2023attentionneed}
\bibfield{author}{\bibinfo{person}{Ashish Vaswani}, \bibinfo{person}{Noam Shazeer}, \bibinfo{person}{Niki Parmar}, \bibinfo{person}{Jakob Uszkoreit}, \bibinfo{person}{Llion Jones}, \bibinfo{person}{Aidan~N. Gomez}, \bibinfo{person}{Lukasz Kaiser}, {and} \bibinfo{person}{Illia Polosukhin}.} \bibinfo{year}{2023}\natexlab{}.
\newblock \bibinfo{title}{Attention Is All You Need}.
\newblock
\newblock
\showeprint[arxiv]{1706.03762}~[cs.CL]
\urldef\tempurl%
\url{https://arxiv.org/abs/1706.03762}
\showURL{%
\tempurl}


\bibitem[Wang et~al\mbox{.}(2022)]%
        {Convex1}
\bibfield{author}{\bibinfo{person}{Yiqiu Wang}, \bibinfo{person}{Rahul Yesantharao}, \bibinfo{person}{Shangdi Yu}, \bibinfo{person}{Laxman Dhulipala}, \bibinfo{person}{Yan Gu}, {and} \bibinfo{person}{Julian Shun}.} \bibinfo{year}{2022}\natexlab{}.
\newblock \bibinfo{title}{ParGeo: A Library for Parallel Computational Geometry}.
\newblock
\newblock
\showeprint[arxiv]{2207.01834}~[cs.CG]
\urldef\tempurl%
\url{https://arxiv.org/abs/2207.01834}
\showURL{%
\tempurl}


\bibitem[{Wikipedia contributors}(2025a)]%
        {wikipedia_hbm}
\bibfield{author}{\bibinfo{person}{{Wikipedia contributors}}.} \bibinfo{year}{2025}\natexlab{a}.
\newblock \bibinfo{title}{High Bandwidth Memory}.
\newblock \bibinfo{howpublished}{\url{https://en.wikipedia.org/wiki/High_Bandwidth_Memory}}.
\newblock
\newblock
\shownote{Accessed: 2025-08-21}.


\bibitem[{Wikipedia contributors}(2025b)]%
        {wikipedia_lpddr}
\bibfield{author}{\bibinfo{person}{{Wikipedia contributors}}.} \bibinfo{year}{2025}\natexlab{b}.
\newblock \bibinfo{title}{LPDDR}.
\newblock \bibinfo{howpublished}{\url{https://en.wikipedia.org/wiki/LPDDR}}.
\newblock
\newblock
\shownote{Accessed: 2025-08-21}.


\bibitem[Williams et~al\mbox{.}(2009)]%
        {10.1145/1498765.1498785}
\bibfield{author}{\bibinfo{person}{Samuel Williams}, \bibinfo{person}{Andrew Waterman}, {and} \bibinfo{person}{David Patterson}.} \bibinfo{year}{2009}\natexlab{}.
\newblock \showarticletitle{Roofline: an insightful visual performance model for multicore architectures}.
\newblock \bibinfo{journal}{\emph{Commun. ACM}} \bibinfo{volume}{52}, \bibinfo{number}{4} (\bibinfo{date}{April} \bibinfo{year}{2009}), \bibinfo{pages}{65–76}.
\newblock
\showISSN{0001-0782}
\urldef\tempurl%
\url{https://doi.org/10.1145/1498765.1498785}
\showDOI{\tempurl}


\bibitem[Wu et~al\mbox{.}(2022)]%
        {wu2022sustainableaienvironmentalimplications}
\bibfield{author}{\bibinfo{person}{Carole-Jean Wu}, \bibinfo{person}{Ramya Raghavendra}, \bibinfo{person}{Udit Gupta}, \bibinfo{person}{Bilge Acun}, \bibinfo{person}{Newsha Ardalani}, \bibinfo{person}{Kiwan Maeng}, \bibinfo{person}{Gloria Chang}, \bibinfo{person}{Fiona~Aga Behram}, \bibinfo{person}{James Huang}, \bibinfo{person}{Charles Bai}, \bibinfo{person}{Michael Gschwind}, \bibinfo{person}{Anurag Gupta}, \bibinfo{person}{Myle Ott}, \bibinfo{person}{Anastasia Melnikov}, \bibinfo{person}{Salvatore Candido}, \bibinfo{person}{David Brooks}, \bibinfo{person}{Geeta Chauhan}, \bibinfo{person}{Benjamin Lee}, \bibinfo{person}{Hsien-Hsin~S. Lee}, \bibinfo{person}{Bugra Akyildiz}, \bibinfo{person}{Maximilian Balandat}, \bibinfo{person}{Joe Spisak}, \bibinfo{person}{Ravi Jain}, \bibinfo{person}{Mike Rabbat}, {and} \bibinfo{person}{Kim Hazelwood}.} \bibinfo{year}{2022}\natexlab{}.
\newblock \bibinfo{title}{Sustainable AI: Environmental Implications, Challenges and Opportunities}.
\newblock
\newblock
\showeprint[arxiv]{2111.00364}~[cs.LG]
\urldef\tempurl%
\url{https://arxiv.org/abs/2111.00364}
\showURL{%
\tempurl}


\bibitem[Wu et~al\mbox{.}(2019)]%
        {accelergy}
\bibfield{author}{\bibinfo{person}{Yannan~Nellie Wu}, \bibinfo{person}{Joel~S Emer}, {and} \bibinfo{person}{Vivienne Sze}.} \bibinfo{year}{2019}\natexlab{}.
\newblock \showarticletitle{Accelergy: An architecture-level energy estimation methodology for accelerator designs}. In \bibinfo{booktitle}{\emph{2019 IEEE/ACM International Conference on Computer-Aided Design (ICCAD)}}.
\newblock


\bibitem[Xiong et~al\mbox{.}(2021)]%
        {mobiledets}
\bibfield{author}{\bibinfo{person}{Yunyang Xiong}, \bibinfo{person}{Hanxiao Liu}, \bibinfo{person}{Suyog Gupta}, \bibinfo{person}{Berkin Akin}, \bibinfo{person}{Gabriel Bender}, \bibinfo{person}{Yongzhe Wang}, \bibinfo{person}{Pieter-Jan Kindermans}, \bibinfo{person}{Mingxing Tan}, \bibinfo{person}{Vikas Singh}, {and} \bibinfo{person}{Bo Chen}.} \bibinfo{year}{2021}\natexlab{}.
\newblock \showarticletitle{Mobiledets: Searching for object detection architectures for mobile accelerators}. In \bibinfo{booktitle}{\emph{Proceedings of the IEEE/CVF conference on computer vision and pattern recognition}}. \bibinfo{pages}{3825--3834}.
\newblock


\bibitem[Xu et~al\mbox{.}(2025)]%
        {10.1145/3695053.3731101}
\bibfield{author}{\bibinfo{person}{Zheng Xu}, \bibinfo{person}{Dehao Kong}, \bibinfo{person}{Jiaxin Liu}, \bibinfo{person}{Jinxi Li}, \bibinfo{person}{Jingxiang Hou}, \bibinfo{person}{Xu Dai}, \bibinfo{person}{Chao Li}, \bibinfo{person}{Shaojun Wei}, \bibinfo{person}{Yang Hu}, {and} \bibinfo{person}{Shouyi Yin}.} \bibinfo{year}{2025}\natexlab{}.
\newblock \showarticletitle{WSC-LLM: Efficient LLM Service and Architecture Co-exploration for Wafer-scale Chips}. In \bibinfo{booktitle}{\emph{Proceedings of the 52nd Annual International Symposium on Computer Architecture}} \emph{(\bibinfo{series}{ISCA '25})}. \bibinfo{publisher}{Association for Computing Machinery}, \bibinfo{address}{New York, NY, USA}, \bibinfo{pages}{1–17}.
\newblock
\showISBNx{9798400712616}
\urldef\tempurl%
\url{https://doi.org/10.1145/3695053.3731101}
\showDOI{\tempurl}


\bibitem[Yan et~al\mbox{.}(2025)]%
        {decoding_sd}
\bibfield{author}{\bibinfo{person}{Minghao Yan}, \bibinfo{person}{Saurabh Agarwal}, {and} \bibinfo{person}{Shivaram Venkataraman}.} \bibinfo{year}{2025}\natexlab{}.
\newblock \showarticletitle{Decoding Speculative Decoding}. In \bibinfo{booktitle}{\emph{Proceedings of the 2025 Conference of the Nations of the Americas Chapter of the Association for Computational Linguistics: Human Language Technologies (Volume 1: Long Papers)}}, \bibfield{editor}{\bibinfo{person}{Luis Chiruzzo}, \bibinfo{person}{Alan Ritter}, {and} \bibinfo{person}{Lu~Wang}} (Eds.). \bibinfo{publisher}{Association for Computational Linguistics}, \bibinfo{address}{Albuquerque, New Mexico}, \bibinfo{pages}{6460--6473}.
\newblock
\showISBNx{979-8-89176-189-6}
\urldef\tempurl%
\url{https://doi.org/10.18653/v1/2025.naacl-long.328}
\showDOI{\tempurl}


\bibitem[Yu et~al\mbox{.}(2022)]%
        {280922}
\bibfield{author}{\bibinfo{person}{Gyeong-In Yu}, \bibinfo{person}{Joo~Seong Jeong}, \bibinfo{person}{Geon-Woo Kim}, \bibinfo{person}{Soojeong Kim}, {and} \bibinfo{person}{Byung-Gon Chun}.} \bibinfo{year}{2022}\natexlab{}.
\newblock \showarticletitle{Orca: A Distributed Serving System for {Transformer-Based} Generative Models}. In \bibinfo{booktitle}{\emph{16th USENIX Symposium on Operating Systems Design and Implementation (OSDI 22)}}. \bibinfo{publisher}{USENIX Association}, \bibinfo{address}{Carlsbad, CA}, \bibinfo{pages}{521--538}.
\newblock
\showISBNx{978-1-939133-28-1}
\urldef\tempurl%
\url{https://www.usenix.org/conference/osdi22/presentation/yu}
\showURL{%
\tempurl}


\bibitem[Yu et~al\mbox{.}(2024)]%
        {Cambricon}
\bibfield{author}{\bibinfo{person}{Zhongkai Yu}, \bibinfo{person}{Shengwen Liang}, \bibinfo{person}{Tianyun Ma}, \bibinfo{person}{Yunke Cai}, \bibinfo{person}{Ziyuan Nan}, \bibinfo{person}{Di Huang}, \bibinfo{person}{Xinkai Song}, \bibinfo{person}{Yifan Hao}, \bibinfo{person}{Jie Zhang}, \bibinfo{person}{Tian Zhi}, \bibinfo{person}{Yongwei Zhao}, \bibinfo{person}{Zidong Du}, \bibinfo{person}{Xing Hu}, \bibinfo{person}{Qi Guo}, {and} \bibinfo{person}{Tianshi Chen}.} \bibinfo{year}{2024}\natexlab{}.
\newblock \showarticletitle{Cambricon-LLM: A Chiplet-Based Hybrid Architecture for On-Device Inference of 70B LLM}. In \bibinfo{booktitle}{\emph{2024 57th IEEE/ACM International Symposium on Microarchitecture (MICRO)}}. \bibinfo{pages}{1474--1488}.
\newblock
\urldef\tempurl%
\url{https://doi.org/10.1109/MICRO61859.2024.00108}
\showDOI{\tempurl}


\bibitem[Zhang et~al\mbox{.}(2024)]%
        {llmcompass}
\bibfield{author}{\bibinfo{person}{Hengrui Zhang}, \bibinfo{person}{August Ning}, \bibinfo{person}{Rohan~Baskar Prabhakar}, {and} \bibinfo{person}{David Wentzlaff}.} \bibinfo{year}{2024}\natexlab{}.
\newblock \showarticletitle{LLMCompass: Enabling Efficient Hardware Design for Large Language Model Inference}. In \bibinfo{booktitle}{\emph{2024 ACM/IEEE 51st Annual International Symposium on Computer Architecture (ISCA)}}. \bibinfo{pages}{1080--1096}.
\newblock
\urldef\tempurl%
\url{https://doi.org/10.1109/ISCA59077.2024.00082}
\showDOI{\tempurl}


\bibitem[Zhao et~al\mbox{.}(2020)]%
        {driving}
\bibfield{author}{\bibinfo{person}{Hengyu Zhao}, \bibinfo{person}{Yubo Zhang}, \bibinfo{person}{Pingfan Meng}, \bibinfo{person}{Hui Shi}, \bibinfo{person}{Li~Erran Li}, \bibinfo{person}{Tiancheng Lou}, {and} \bibinfo{person}{Jishen Zhao}.} \bibinfo{year}{2020}\natexlab{}.
\newblock \showarticletitle{Driving Scenario Perception-Aware Computing System Design in Autonomous Vehicles}. In \bibinfo{booktitle}{\emph{2020 IEEE 38th International Conference on Computer Design (ICCD)}}. \bibinfo{pages}{88--95}.
\newblock
\urldef\tempurl%
\url{https://doi.org/10.1109/ICCD50377.2020.00031}
\showDOI{\tempurl}


\bibitem[Zheng et~al\mbox{.}(2023a)]%
        {tileflow}
\bibfield{author}{\bibinfo{person}{Size Zheng}, \bibinfo{person}{Siyuan Chen}, \bibinfo{person}{Siyuan Gao}, \bibinfo{person}{Liancheng Jia}, \bibinfo{person}{Guangyu Sun}, \bibinfo{person}{Runsheng Wang}, {and} \bibinfo{person}{Yun Liang}.} \bibinfo{year}{2023}\natexlab{a}.
\newblock \showarticletitle{TileFlow: {A} Framework for Modeling Fusion Dataflow via Tree-based Analysis}. In \bibinfo{booktitle}{\emph{Proceedings of the 56th Annual {IEEE/ACM} International Symposium on Microarchitecture, {MICRO} 2023, Toronto, ON, Canada, 28 October 2023 - 1 November 2023}}. \bibinfo{publisher}{{ACM}}, \bibinfo{pages}{1271--1288}.
\newblock
\urldef\tempurl%
\url{https://doi.org/10.1145/3613424.3623792}
\showDOI{\tempurl}


\bibitem[Zheng et~al\mbox{.}(2023b)]%
        {chimera}
\bibfield{author}{\bibinfo{person}{Size Zheng}, \bibinfo{person}{Siyuan Chen}, \bibinfo{person}{Peidi Song}, \bibinfo{person}{Renze Chen}, \bibinfo{person}{Xiuhong Li}, \bibinfo{person}{Shengen Yan}, \bibinfo{person}{Dahua Lin}, \bibinfo{person}{Jingwen Leng}, {and} \bibinfo{person}{Yun Liang}.} \bibinfo{year}{2023}\natexlab{b}.
\newblock \showarticletitle{Chimera: An Analytical Optimizing Framework for Effective Compute-intensive Operators Fusion}. In \bibinfo{booktitle}{\emph{2023 IEEE International Symposium on High-Performance Computer Architecture (HPCA)}}. \bibinfo{pages}{1113--1126}.
\newblock
\urldef\tempurl%
\url{https://doi.org/10.1109/HPCA56546.2023.10071018}
\showDOI{\tempurl}


\bibitem[Zhong et~al\mbox{.}(2024)]%
        {distservedisaggregatingprefilldecoding}
\bibfield{author}{\bibinfo{person}{Yinmin Zhong}, \bibinfo{person}{Shengyu Liu}, \bibinfo{person}{Junda Chen}, \bibinfo{person}{Jianbo Hu}, \bibinfo{person}{Yibo Zhu}, \bibinfo{person}{Xuanzhe Liu}, \bibinfo{person}{Xin Jin}, {and} \bibinfo{person}{Hao Zhang}.} \bibinfo{year}{2024}\natexlab{}.
\newblock \bibinfo{title}{DistServe: Disaggregating Prefill and Decoding for Goodput-optimized Large Language Model Serving}.
\newblock
\newblock
\showeprint[arxiv]{2401.09670}~[cs.DC]
\urldef\tempurl%
\url{https://arxiv.org/abs/2401.09670}
\showURL{%
\tempurl}


\end{thebibliography}

%%%%%%%%%%%%%%%%%%%%%%%%%%%%%%%%%%%%

\end{document}